\documentclass[useAMS,usenatbib]{mnras}
\usepackage{graphicx}
\usepackage{longtable}
\usepackage{amssymb}
\usepackage{amsmath}
\title[X-ray emission from high-z galaxies]{The deepest X-ray view of high-redshift galaxies: constraints on low-rate black-hole accretion}
\author[F. Vito et al.]
{F. Vito$^{1,2,3,4}$\thanks{E-mail: fvito@psu.edu},
R. Gilli$^{3}$,
C. Vignali$^{3,4}$,
W.N. Brandt$^{1,2,5}$,
A. Comastri$^{3}$, 
\newauthor
G. Yang$^{1,2}$,
B.D. Lehmer$^{6}$,
B. Luo$^{7}$,
A. Basu-Zych$^{8,9}$,
F.E. Bauer$^{10,11,12,13}$,
\newauthor
N. Cappelluti$^{14,15}$,
A. Koekemoer$^{16}$, 
V. Mainieri$^{17}$, 
M. Paolillo$^{18,19,20}$,
P. Ranalli$^{21}$,
\newauthor
O. Shemmer$^{22}$,
J. Trump$^{1,23}$,
J.X. Wang$^{24}$, 
Y.Q. Xue$^{24}$
\\ \\
$^{1}$ Department of Astronomy \& Astrophysics, 525 Davey Lab, The Pennsylvania State University, University Park, PA 16802, USA\\
$^{2}$ Institute for Gravitation and the Cosmos, The Pennsylvania State University, University Park, PA 16802, USA\\
$^{3}$ INAF -- Osservatorio Astronomico di Bologna, Via Ranzani 1, 40127 Bologna, Italy\\
$^{4}$ Dipartimento di Fisica e Astronomia, Universit\`a degli Studi di Bologna, Viale Berti Pichat 6/2, 40127 Bologna, Italy \\
$^{5}$ Department of Physics, The Pennsylvania State University, University Park, PA 16802, USA\\
$^{6}$ Department of Physics, University of Arkansas, 226 Physics Building, 835 West Dickinson Street, Fayetteville, AR 72701, USA\\
$^{7}$ School of Astronomy and Space Science, Nanjing University, Nanjing 210093, China\\ 
$^{8}$ NASA Goddard Space Flight Center, Code 662, Greenbelt, MD 20771\\
$^{9}$ Department of Physics, University of Maryland Baltimore County, Baltimore, MD 21250, USA\\
$^{10}$ Instituto de Astrof\'isica, Facultad de F\'isica, Pontificia Universidad Cat\'olica de Chile, Casilla 306, Santiago 22, Chile\\
$^{11}$ EMBIGGEN Anillo, Chile\\
$^{12}$ Millennium Institute of Astrophysics, Chile\\
$^{13}$ Space Science Institute, 4750 Walnut Street, Suite 205, Boulder, CO 80301, USA\\
$^{14}$ Department of Physics, Yale University, P.O. Box 208121, New Haven, CT 06520, USA\\
$^{15}$ Yale Center for Astronomy \& Astrophysics, Physics Department, P.O. Box 208120, New Haven, CT 06520, USA\\
$^{16}$ Space Telescope Science Institute 3700 San Martin Drive, Baltimore MD 21218, USA \\
$^{17}$ European Southern Observatory, Karl-Schwarzschild-str. 2, 85748 Garching bei Munchen, Germany\\
$^{18}$ Dip.di Fisica Ettore Pancini, University of Naples ``Federico II'', C.U. Monte SantAngelo, Via Cinthia, 80126, Naples, Italy\\
$^{19}$ INFN Sezione di Napoli, Via Cinthia, I-80126 Napoli, Italy\\
$^{20}$ Agenzia Spaziale Italiana - Science Data Center, Via del Politecnico snc, 00133 Roma, Italy\\
$^{21}$ Lund Observatory, Box 43, 22100 Lund, Sweden \\
$^{22}$ Department of Physics, University of North Texas, Denton, TX 76203\\
$^{23}$ Hubble Fellow\\
 $^{24}$ CAS Key Laboratory for Research in Galaxies and Cosmology, Department of Astronomy, \\ University of Science and Technology of China, Hefei, Anhui 230026, China\\
}

\newcommand*{\unit}[1]{~\mathrm{#1}}

\newcommand*{\chandra}{\textit{Chandra}}

\newcommand*{\chapt}[1]{\S~\ref{#1}}
\newcommand*{\Hband}{\textit{H}-band\,}
\newcommand{\angstrom}{\mbox{\normalfont\AA}}

\newcommand*{\srcstack}{$\mathrm src_{stack}$\,}
\newcommand*{\srcxdet}{$\mathrm src_{det}$\,}

\begin{document}

\date{}

\graphicspath{{.}}
% \date{}
%\pagerange{\pageref{firstpage}--\pageref{lastpage}} \pubyear{2014}

\maketitle
%  \label{firstpage}

\begin{abstract}
We exploit the 7 Ms \textit{Chandra} observations in the \chandra\,Deep Field-South (\mbox{CDF-S}), the deepest X-ray survey to date, coupled with CANDELS/GOODS-S data, to measure the total X-ray emission arising from 2076 galaxies at $3.5\leq z < 6.5$. This aim is achieved by stacking the \textit{Chandra} data at the positions of optically selected galaxies, reaching effective exposure times of $\geq10^9\rmn{s}$. We detect significant (\mbox{$>3.7\sigma$}) X-ray emission from massive galaxies at $z\approx4$. We also report the detection of massive galaxies at $z\approx5$ at a $99.7\%$ confidence level ($2.7\sigma$), the highest significance ever obtained for X-ray emission from galaxies at such high redshifts. No significant signal is detected from galaxies at even higher redshifts. The stacking results place constraints on the BHAD associated with the known high-redshift galaxy samples, as well as on the SFRD at high redshift, assuming a range of prescriptions for X-ray emission due to X-
ray 
binaries. We find that the X-ray emission from our sample is likely dominated by processes related to 
star formation. Our results 
show that low-rate mass accretion 
onto SMBHs in individually X-ray-undetected galaxies is negligible, compared with the BHAD measured for samples of X-ray detected AGN, for cosmic SMBH mass assembly at high redshift. We also place, for the first time, constraints on the faint-end of the AGN X-ray luminosity function ($\rmn{logL_X\sim42}$) at $z>4$, with evidence for fairly flat slopes. The implications of all of these findings are discussed in the context of the evolution of the AGN population at high redshift. 
\end{abstract}


\section{Introduction}
Several empirical relations exist between the masses of Super-Massive Black Holes (SMBHs), which are thought to reside in the centres of nearly all galaxies, and the properties of the host galaxies themselves, such as mass, luminosity and velocity dispersion of the bulge (e.g. \citealt{Magorrian98}; \citealt{Ferrarese00}, \citealt{Gebhardt00}; \citealt{Marconi03}, \citealt{Gultekin09}). Such relations may be the results of an evolutionary path that SMBHs and host galaxies share, despite their very different physical scales. This so-called ``BH-galaxy co-evolution" is also suggested by the similar redshift evolution of the cosmic Star Formation Rate Density (SFRD) and BH Accretion Rate Density (BHAD): both quantities peak at $z\sim1-3$ and decrease toward lower and higher redshift (e.g. \citealt{Boyle98}; \citealt{Marconi04}; \citealt{Hopkins06};  \citealt{Aird10,Aird15}; \citealt{Delvecchio14}). Moreover, both galaxies and Active Galactic Nuclei (AGN, which trace 
the nuclear accretion and hence the SMBH mass growth) are characterized by a ``downsizing" evolution, in the sense that the build-up of massive galaxies and the peak of the space density of luminous AGN occurred at earlier cosmic times ($z\sim2-3$) than those of less massive galaxies and less luminous AGN (e.g. \citealt{Cowie96};  \citealt{Barger05}; \citealt{Hasinger05}; \citealt{LaFranca05}; \citealt{Thomas05}; \citealt{Damen09}; \citealt{Ebrero09}, \citealt{Yencho09}, \citealt{Aird10}; \citealt{Delvecchio14}; \citealt{Ueda14}). This interplay between SMBH and galaxy (see \citealt{Brandt15}, and references therein) is expected to be crucial during the phases of galaxy formation in the Early Universe. However, while SFRD evolution has been studied up to $z\sim10$ \citep[e.g.][]{Bouwens15}, our knowledge of AGN evolution is largely limited to later cosmic times. In fact, despite the discovery of an increasing number of 
luminous ($\rmn{L_{bol}} 
> 10^{13} \rmn{L_\odot}$) $z>6$ quasars in the last 15 years \citep[e.g.][]{Willott03, Willott09, Fan06a, Fan06b, Jiang08, Jiang09, Mortlock11, Venemans13, Venemans15a, Venemans15b, Banados14, Wu15}, the bulk of the AGN population at high redshift (i.e. low-to-moderate luminosity and possibly obscured AGN) is still largely unknown.

Besides the issue of galaxy formation and evolution, shedding light on the population of low-to-moderate luminosity ($\rmn{L_{bol}}=10^{43-44}\rmn{erg\,s^{-1}}$) AGN at $z\gtrsim4$ is also mandatory to unveil SMBH assembly processes in the early Universe. The two most promising classes of models are usually referred to as ``light seeds" models, whereby the primordial black holes have masses of $\sim10^{2}\rmn{M_{\odot}}$ and are the remnants of Pop III stars, and ``heavy seeds" models, which ascribe the formation of black holes with masses $10^{4-5}\rmn{M_\odot}$ to the direct collapse of massive clouds of pristine gas (see \citealt{Volonteri10} and \citealt{Haiman13} and references therein). The AGN luminosity functions predicted by different seeding models are indistinguishable at low redshift (where the BH occupation fraction has been recently used to test seeding models, with inconclusive results; e.g. \citealt{Miller15, Trump15}) but start to diverge at $z\gtrsim4$, particularly with regard to the 
evolution of the faint end of the AGN luminosity function. 

While wide-field optical/IR surveys (e.g. SDSS, UKIDSS) coupled with dedicated spectroscopic campaigns \citep[e.g.][]{Mortlock11, Venemans13,Venemans15a}, have been successfully used to select high-redshift luminous ($\rmn{log}L_{bol}>47$) QSOs, AGN with luminosities below the break of the luminosity function can currently be detected only by deep X-ray surveys. In fact, X-rays are much less affected by galaxy dilution and obscuration than the optical/near-IR bands \citep[e.g.][]{Brandt15}. This is particularly important when searching for low-to-moderate luminosity ($\rmn{log}L_{X}\lesssim 44$) AGN at high redshift as emission from the host galaxy can outshine the faint nuclear radiation in the optical band. Moreover, many models of SMBH growth predict that most of the mass assembly in the early Universe occurs during phases of strong obscuration, when nuclear emission is completely hidden in the optical/IR bands \citep[e.g.][]{Hopkins08}. Hence it is not surprising that many authors in recent years focused 
on the $3\lesssim z\lesssim5$ range and tried to 
assess the 
evolution of high-redshift AGN 
exploiting data from X-ray surveys like the XMM-\textit{Newton} \citep{Brusa09} and \textit{Chandra} \citep[Marchesi et al. submitted]{Civano11} observations in the COSMOS field, the 4~Ms \textit{Chandra} Deep Field South \citep[CDF-S;][]{Vito13}, the Subaru-XMM Deep Survey \citep[SXDS;][]{Hiroi12} or from a combination of surveys \citep{Kalfountzou14, Vito14, Georgakakis15}. All of these works found that the space density of luminous ($\rmn{logL_X}\gtrsim44$) AGN declines exponentially by a factor of $\sim10$ from $z=3$ to 5. However, even the works exploiting the very deep 4~Ms CDF-S data \citep{Xue11} could not collect enough low-luminosity AGN to assess the evolution of the faint end of the luminosity function at $z>3$. In particular, \cite{Vito14} and \cite{Georgakakis15} found that a Pure Density Evolution (PDE) model is the best representation of the AGN luminosity function at $z>3$, although the data in these works were insufficient to disentangle such a model from a more complex 
Luminosity-Dependent Density Evolution (LDDE) model, which is often invoked for the low-redshift AGN population (e.g. \citealt{Miyaji00, Miyaji15}; \citealt{Ueda03,Ueda14}; \citealt{Hasinger05}; \citealt{LaFranca05}; \citealt{Silverman08}; \citealt{Ebrero09}; \citealt{Yencho09}; \citealt{Fotopoulou16}; but see also \citealt{Aird10,Aird15,Buchner15,Ranalli16}). Moreover, the lack of a significant number of X-ray detected AGN at $z\gtrsim5$ \citep{Vignali02, Steffen04, Barger05,Civano12, Weigel15}, poses a high-redshift limit to those works.

Stacking the X-ray images at the positions of known galaxies can be used to efficiently increase the sensitivity by large factors and thereby to investigate the average properties of the population of low-luminosity AGN, that are individually undetected even in the deepest surveys. In previous works, \cite{Cowie12}, \cite{Basu-Zych13} and \cite{Treister13} did not detect significant X-ray signal at $z\gtrsim5$ by applying their stacking procedures to the deep 4~Ms CDF-S data. \cite{Cowie12}, in particular, concluded that the X-ray emission they retrieved from high-redshift galaxies was consistent with pure stellar processes.
The increasing contribution from star-forming galaxies to the total X-ray emission at the lowest fluxes probed by deep surveys \citep{Lehmer12} poses a challenge to the studies of low-luminosity AGN: high- and low-mass X-ray binaries in galaxies can produce a luminosity up to $L_X\sim10^{42}\rmn{erg\,s^{-1}}$ \citep[e.g.][]{Ranalli03,Ranalli05}. On the one hand 
this 
contribution is treated as 
contaminating emission by the works aimed at studying AGN, but on the other hand it could be used to derive relations between the X-ray luminosity and the Star Formation Rate (SFR) of a galaxy (e.g. \citealt{Bauer02}, \citealt{Ranalli03}, \citealt{Basu-Zych13_2}, \citealt{Symeonidis14}, Lehmer et al., accepted), and hence place important constraints on galaxy evolution at high redshift. 

Stacking procedures with deep X-ray data rely on accurate knowledge of the positions and redshifts for a large number of objects of interest: this typically requires a combination of deep X-ray and optical/near-IR observations. The 7 Ms CDF-S (Luo et al., in prep.) is the deepest X-ray survey to date, reaching a flux limit of $F_{0.5-2\,\rmn{keV}}\sim4.5\times10^{-18}\rmn{erg\,cm^{-2}s^{-1}}$, and is one of the fields with the most-complete and deepest multiwavelength coverage in the sky, e.g. it is part of the Great Observatories Origins Deep Survey \citep[GOODS-S;][]{Dickinson03}. In particular, it was observed with the near-infrared WFC3 on board the Hubble Space Telescope (\textit{HST}) in the Cosmic Assembly Near-Infrared Deep Extragalactic Legacy Survey \citep[CANDELS;][]{Grogin11, Koekemoer11} plus the Hubble Ultra-Deep Field \citep[UDF;][]{Beckwith06}, reaching a depth of $m_{AB}\sim28-30$ in the \Hband \citep{Guo13}. Thanks to the unprecedented multiwavelength coverage of CANDELS/
GOODS-S 
at faint magnitudes and to the superb Hubble spatial resolution, the CANDELS team 
compiled a high-quality spectroscopic and photometric redshift catalog \citep{Dahlen13, Santini15}.

In this work we exploit the data from the 7 Ms CDF-S and perform a stacking analysis of samples of CANDELS-selected galaxies at $3.5\leq z<6.5$. Our purpose is to place constraints on the BHAD and on the faint end of the AGN luminosity function at high redshift. Moreover we will use the stacked X-ray signal to derive information on the star-formation properties of high-redshift galaxies. Throughout this paper we will use a $H_0=70\,\rmn{km\,s^{-1}Mpc^{-1}}$, $\Omega_m=0.3$, and $\Omega_\Lambda=0.7$ cosmology. Errors and upper limits are quoted at the $1\sigma$ confidence level, unless otherwise noted. One \textit{Chandra} pixel is $0.492''\times0.492''$.

\section{The stacking procedure}\label{sec1}

\begin{figure}
\centering
\includegraphics[width=80mm,keepaspectratio]{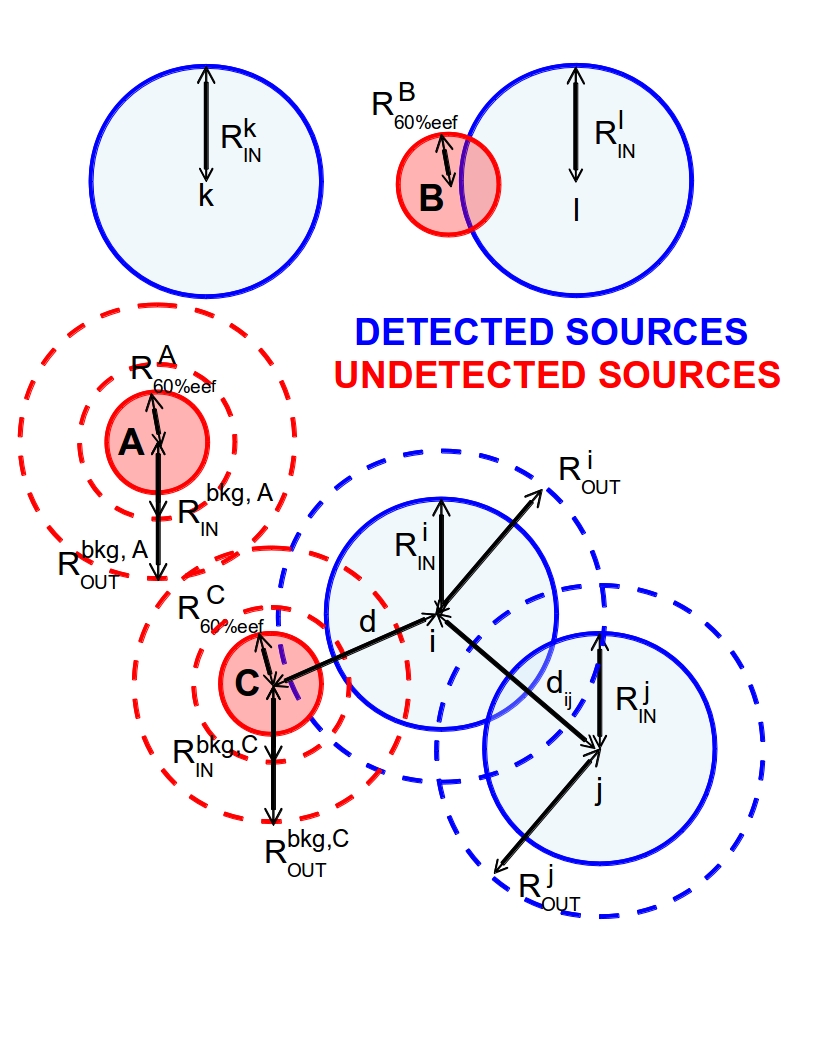}
\caption{ Example of possible configuration of sources to be stacked ({\it A}, {\it B} and {\it C}, in red) with respect to X-ray detected sources ({\it i}, {\it j}, {\it k} and {\it l}, in blue). We assumed an off-axis angle of $\theta\sim5'$, and therefore extraction regions corresponding to the 60\% EEF are used. Source {\it A} represents the simplest case of an isolated, undetected source whose background annulus region is not affected by any nearby individually-detected source. Source {\it B} overlaps with an individually-detected source. The background of source {\it C} is partially contaminated by the X-ray emission of the detected source $i$. 
Following the procedure in \chapt{sec1}, {\it A} and {\it C} will be stacked while {\it B} is rejected and will not be stacked. All the X-ray detected sources $i$, $j$ and $l$ are masked. Note that source $j$ is excluded when the background used to refill the masked region corresponding to source $i$ is evaluated.}
\label{fig1}
\end{figure}

Here we briefly describe the main steps of the aperture photometry procedure used to estimate the net count rate at given positions on the X-ray image (see Fig.~\ref{fig1} for a visual representation): 

\begin{enumerate}
\item  Through detailed simulations of the \chandra\, observations in the CDF-S, we derived a parametrization of the \chandra\, Point Spread Function (PSF) as a function of the off-axis angle $\theta$ (see Appendix~\ref{psf}). 
 \item we compute the radius corresponding to the 90\% Encircled Energy Fraction (EEF; at 1.5 keV) at the position of each source to be stacked (\srcstack), $\mathrm{R_{90\%EEF}^{\mathrm{stack}}}$, and each detected X-ray source (\srcxdet), $\mathrm{R_{90\%EEF}^{\mathrm{det}}}$. 
 
 \item For each \srcstack, we define an annulus region in which the background of the \srcstack photometry is evaluated centred at the position to be stacked with inner radius  $\mathrm{R_{IN}^{bkg_{stack}}=1.1\times R_{90\%EEF}^{\mathrm{stack}}}$ and outer radius $\mathrm{R_{OUT}^{bkg_{stack}}=R_{IN}^{bkg_{stack}} + 20\,pixels}$.\footnote{Different choices of the inner radius factor (1.0-1.2) and annulus width (10-30 pixels) do not affect significantly the final results.} Although the background is sampled better at large $\theta$ as the annulus widths are fixed (to 20 pixels), the number of photons detected at positions close to the aim point is sufficient to estimate the background well. While, \textit{on average}, there is no offset between the optical and X-ray positions in the 7 Ms CDF-S (see \chapt{stacking}), the \textit{individual} offsets are spread around zero. Therefore, stacking the X-ray emission at the optical positions would return a lower EEF than the expected one at a given $\theta$. 
To 
account for this effect statistically, in Appendix~\ref{psf} we include the spread of the X-ray-optical position offsets in the simulations of the \chandra\, PSF in the CDF-S. When masking the detected X-ray sources we use the nominal PSFs, while when stacking the optical positions we used the PSFs corrected for the statistical offset of the positions.
 
  \item For each \srcxdet, we define a background annulus with inner radius $\mathrm{R_{IN}^{bkg_{det}}=n\times R_{90\%EEF}^{det}}$, where $n=1.5$, 1.75, and 2 for \srcxdet with $<100$, $100-1000$, and $>1000$ net counts, respectively, and outer radius $\mathrm{R_{OUT}^{bkg_{det}}=R_{IN}^{bkg_{det}}+20\,pixels}$.
 
  \item We masked all of the \srcxdet. For every \srcxdet, the masked region has a radius of $\mathrm{R_{IN}^{bkg_{det}}}$. This radius is large enough to mask all the emission of even the brightest X-ray sources. The background used to refill the masked region is evaluated from the background annulus of the detected source (defined in the previous point). For any $\mathrm src_{det}^i$ to be masked, the procedure checks that every other $\mathrm src_{det}^j$ is located at a distance $d_{ij}>\mathrm{R_{OUT}^{bkg_{det},i}}+\mathrm{R_{IN}^{bkg_{det},j}}$, otherwise $\mathrm src_{det}^j$ is also masked to prevent contamination in the annulus region where the background to refill the masked $\mathrm src_{det}^i$ area is evaluated.

 \item For each \srcstack, we defined the radius of the circular region used to compute the source photometry as the radius corresponding to the 80\%, 75\%, 60\%, and 40\% EEF for $\theta<3.5'$, $3.5'<\theta<4.25'$, $4.25'<\theta<5.5'$, and $5.5'<\theta<7.8'$, respectively.
We motivate this choice through simulations described in \chapt{sec2} below.

  \item For each \srcstack, we compute the distance $d$ to every \srcxdet. If $d\leq R^{phot} + \mathrm{R_{IN}^{bkg_{det}}}$, \srcstack would fall in a masked region and it is therefore excluded from the stacking analysis. The fraction of CANDELS area lost because of the presence of an X-ray detected source (and therefore excluded by the stacking procedure) is $\sim0.10$. 
  
 \item The net photometry of each \srcstack is computed by rescaling the background count rate by the ratio of the source and background region areas and subtracting it from the count rate measured in the source region. The net count rate of each \srcstack is divided by an aperture-correction factor $P(\theta)=$0.80, 0.75, 0.60, and 0.40 for sources at $\theta<3.5'$, $3.5'<\theta<4.25'$, $4.25'<\theta<5.5'$, and $5.5'<\theta<7.8'$, respectively, in order to obtain the net count rate corresponding to the 100\% EEF. 
 
 \item The total (i.e. summed over all \srcstack) count rate and errors are computed through a bootstrapping procedure: given a list of {\it N} positions to be stacked and the relative net count rates, we constructed 1000 new lists by randomly picking up a number {\it N} of net count rates from the original list, allowing for repetition. We therefore derive a distribution of 1000 total count rates. Its mean and $1\sigma$ scatter is defined to be the final total net count rate and $1\sigma$ uncertainty. 
 \end{enumerate}
 
\subsection{Validation of the procedure through simulations}\label{sec2}

To check the stacking procedure, we applied it to a simulated image of the CDF-S. In order to keep the required computation time reasonably low, the checking is based on only the first 54 \textit{Chandra} observations, for a total of $\sim4\, \rmn{Ms}$ \citep{Xue11}. While the \textit{Chandra} PSF is known to vary along both the radial and azimuthal directions, the similar aim points and different roll angles of the \textit{Chandra} observations in the CDF-S smear out the azimuthal variations of the PSF in different pointings. In practice, the final PSF at a given point on the field is fairly well determined by only its angular separation from the average CDF-S aim point, which does not change significantly with the inclusion of the most recent \textit{Chandra} pointings.

\begin{figure*}
\centering
\includegraphics[width=160mm,keepaspectratio]{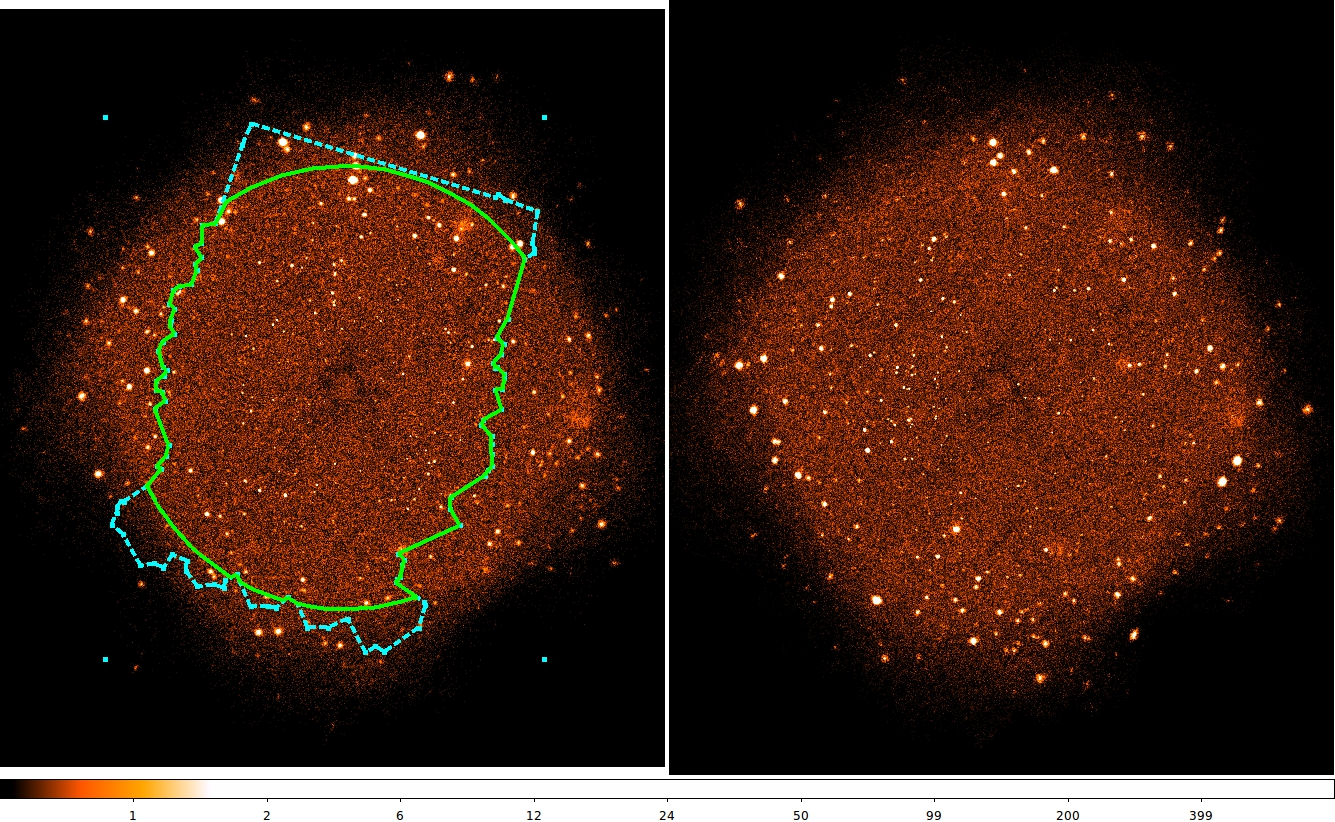}
\caption{Real (left panel) and simulated (right panel) merged and smoothed soft-band images of the CDF-S. The dashed cyan region shows the CANDELS coverage of the field, and the solid green region combines the CANDELS coverage with a restriction to Chandra observations of $\theta < 7.8\arcmin$. }
\label{fig4}
\end{figure*}

We simulated a mock catalog of AGN whose soft-band flux distribution follows the logN-logS relation from \cite{Gilli07} and a clustering recipe was utilized to produce the mock positions. Technical details on the simulations are given in Appendix~\ref{psf}. To reproduce the real case, we merged the final image of simulated sources with the soft-band background map, derived by masking all the X-ray detected sources \citep{Xue11} in the real merged soft-band 4~Ms CDF-S image following the procedure described in \chapt{sec1}. The output is a 4~Ms-like soft-band image of mock sources with real background (see Fig.~\ref{fig4} for a comparison with the real image). We run {\it wavdetect} with a threshold of $10^{-6}$ to create a mock catalog of detected sources and we derived the net counts for each undetected source following the above-described 
procedure and using the fitted radii corresponding to 
different EEF, including the statistical effect of X-ray-optical offsets (see Appendix~\ref{psf}), for the aperture photometry.
 We restricted the field of investigation to the area within $7.8'$ from the 4~Ms CDF-S average aim point because at larger off-axis angles, where the \textit{Chandra} sensitivity drops and a smaller number of pointings overlap, the sharp gradients in the exposure caused issues during the evaluation of the background.\footnote{ In presence of strong exposure gradients, the background counts in a circular region is not always well described by simply rescaling the background counts in a surrounding annulus region.}  We apply the same off-axis angle cut to the real data in \chapt{sec3.1}. 
 
  The simulation software records the number of detected X-ray photons for all the simulated sources, including those undetected by {\it wavdetect}. We checked the accuracy of the stacking code by applying the aperture photometry procedure to each undetected simulated source and comparing the number of retrieved photons with that recorded by the simulation software for that very source, applying the proper EEF correction.
 We found an overall agreement within $\sim10\%$ uncertainty.

\begin{figure}
\centering
\includegraphics[width=80mm,keepaspectratio]{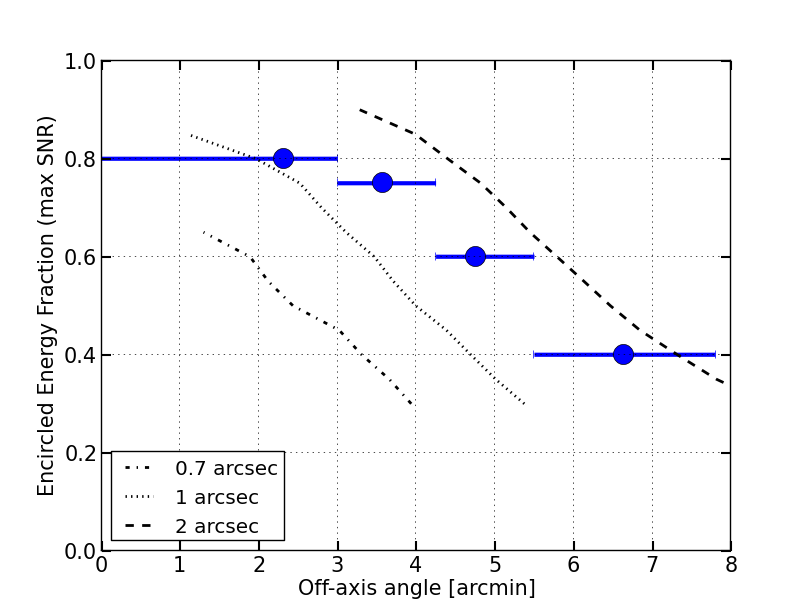}
\caption{Encircled Energy Fraction corresponding to the maximum SNR, as derived from the simulations discussed in \chapt{sec2}, as a function of $\theta$. Three curves of constant angular dimension are also shown for reference.}
\label{SNR}
\end{figure}

 We have also used the photometry of the simulated image to check which EEF returns the highest output signal-to-noise ratio (SNR), computed as the total count rate divided by its $1\sigma$ dispersion as derived from the bootstrap procedure (see end of \chapt{sec1}), as a function of $\theta$. We considered four $\theta$ bins, chosen to include similar numbers of real CANDELS galaxies. Fig.~\ref{SNR} shows the EEF returning the maximum SNR in each $\theta$ bin. We therefore chose to use the radii corresponding to these EEF at different $\theta$ to perform aperture photometry of real sources. For reference, Fig.~\ref{SNR} also reports three curves corresponding to constant radii of $0.7''$, $1''$ and $2''$. We note that, while the EEF corresponding to the maximum SNR decreases at increasing off-axis angles, the angular dimension of the extraction radii increases from $\sim0.8''$ at $\theta=0'$ to $2''$ at $\theta=7.8'$. This behaviour was already pointed out in previous works \citep[e.g.,][]{Cowie12}, 
and 
it is due to the non-trivial radial variations of the \chandra\, PSF, effective area and background.

\section{Stacking CANDELS galaxies at $3.5\leq z <6.5$}\label{sec3}
\subsection{The samples}\label{sec3.1}
We selected three samples of galaxies with $\theta\le 7.8'$ from the 7 Ms CDF-S average aim point from the \mbox{GOODS-S/CANDELS} catalog \citep{Guo13, Santini15} in different redshift bins at $z\geq3.5$ using 
both spectroscopic (available for only $\sim2\%$ of the galaxies in the final samples) and photometric redshifts as provided by \cite{Santini15}.
Photometric redshifts at high redshift are generally fairly accurate, as the Lyman-break, one of the most prominent spectral features in galaxies, is redshifted into the optical bands and is recognizable in the observed galaxy Spectral Energy Distribution (SED). Moreover, the CANDELS photometric dataset is the deepest available and photometric redshifts based on that are expected to be very accurate \citep{Santini15}. Nonetheless, in Appendix~\ref{dz} we investigate photometric-redshift accuracies as a function of redshift, and in Appendix~\ref{PDF} we assess the impact of redshift uncertainties on the stacking results. 
                                                                                                                          
\begin{figure}
% \centering
\includegraphics[width=80mm,keepaspectratio]{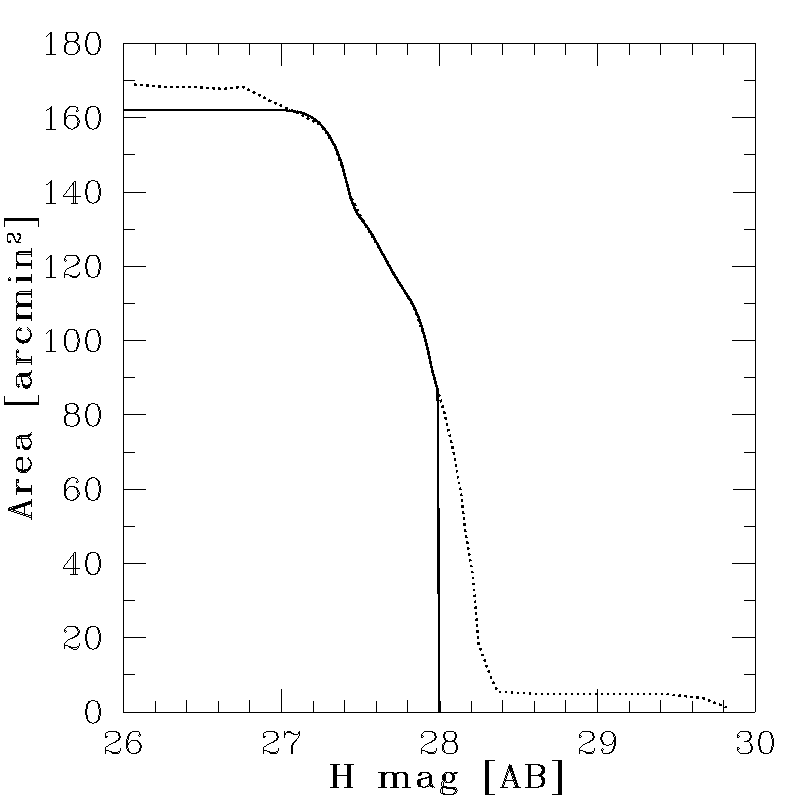}
\caption{CANDELS sky-coverage as a function of the \Hband magnitude used in this work (solid line), resulting from the original CANDELS sky-coverage (dotted line, \citealt{Guo13}) applying the cut in off-axis angle ($\theta<7.8'$) and magnitude (H$<28$), as described in \chapt{sec3.1}.}%

\label{fig8}
\end{figure}

\begin{figure*}
\centering
\includegraphics[width=160mm,keepaspectratio]{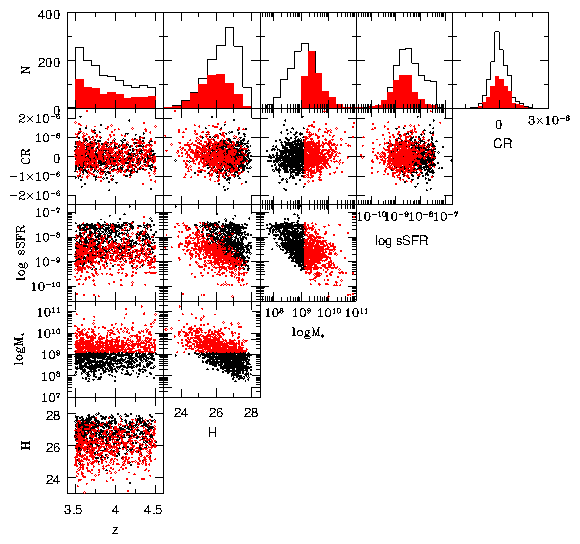}
\caption{Distributions of redshift, \Hband magnitude, stellar mass, sSFR and soft-band net count-rate for all the galaxies stacked in the $3.5\leq z<4.5$ sample. Black points and histograms show the entire sample, while red points and histograms represent the most massive half of galaxies, defined as in Tab.~\ref{tab2}.}
\label{par1}
\end{figure*}

\begin{figure*}
\centering
\includegraphics[width=160mm,keepaspectratio]{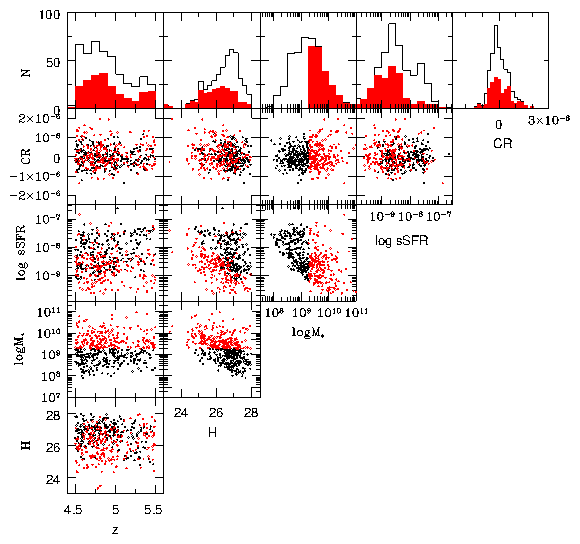}
\caption{Same as Fig.~\ref{par1}, but for the $4.5\leq z<5.5$ sample.}
\label{par2}
\end{figure*}

\begin{figure*}
\centering
\includegraphics[width=160mm,keepaspectratio]{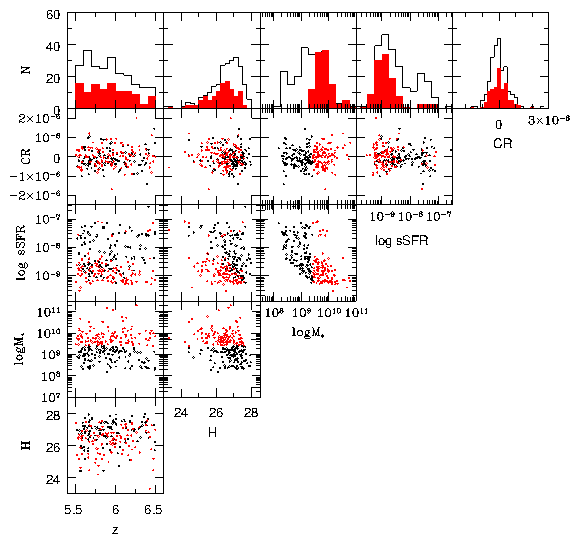}
\caption{Same as Fig.~\ref{par1}, but for the $5.5\leq z<6.5$ sample.}
\label{par3}
\end{figure*}

 \cite{Santini15} provide estimates of $M_*$ and SFR for every CANDELS galaxy. In particular, multiple estimates of SFR were computed for each object under different physical assumptions (e.g. initial mass functions, star formation histories, etc.) and using different fitting methods and codes. In this work, we used the estimates flagged as {\it SFR\_2a\_tau} in \cite{Santini15}, which were derived assuming a Chabrier IMF and an exponentially decreasing SFH. The mean SFR and $M_*$ and relative $1\sigma$ uncertainties for each sample were derived applying a bootstrap procedure.
The distributions of redshift, magnitude, stellar mass, specific Star Formation Rate ($\rmn{sSFR=SFR/M_* }$) and net X-ray count rate (from the stacking procedure) for all the individual galaxies are shown in Fig.~\ref{par1}, \ref{par2} and \ref{par3} for samples at $3.5\leq z<4.5$, $4.5\leq z<5.5$ and $5.5\leq z<6.5$, respectively. 
Tab.~\ref{tab2} reports the main characteristics of the considered samples. For each redshift interval, we stacked both all the galaxies and only the most massive ones. The mass cut we applied in the latter case corresponds to the median mass in each redshift interval as reported in Tab.~\ref{tab2}. 
The CANDELS sky-coverage as a 
function of H magnitude is reported in \cite{Guo13}. Rejecting the area at $\theta>7.8'$ resulted in an area loss of $\sim 10 \unit{armin^2}$. The CANDELS field used in this work (green region) is compared to the original one (blue region) in Fig.~\ref{fig4}. Using the resulting sky-coverage and the H magnitudes provided by \cite{Guo13}, we assigned each galaxy $i$ a weight factor $\Omega_i=\rmn{\frac{162}{coverage(H_i)}}\geq1$, to properly account for the incompleteness due to the smaller sensitive CANDELS area at faint magnitudes, where 162 arcmin$^2$ is the maximum (i.e. for bright magnitudes) field covered by CANDELS in the CDF-S (see Fig.~\ref{fig8}). We applied a cut in magnitude to the sample, excluding from the stacking analysis galaxies with $\rmn{H}\geq28$, as these few galaxies would dominate the results due to their large weighting factors $\Omega_i$. The final CANDELS sky-coverage (rescaled and cut at $H<28$) is shown in Fig.~\ref{fig8}. No further cuts apart from those in magnitude and $\theta$ 
were applied to the samples of CANDELS galaxies. The magnitude threshold we used, $H=28$, corresponds at $z>3.5$ to a stellar mass of a few $\times10^8\,\rmn{M_\odot}$ (see Fig.~\ref{par1}), a factor of $\sim10$ lower than the Large Magellanic Cloud mass \citep[e.g.][]{vanderMarel02}. While a large number of galaxies are expected to have magnitudes fainter than that, unless most of the BH growth occurred in $<10^8\,\rmn{M_\odot}$ galaxies, there is no reason to believe that their exclusion from the stacking analysis would significantly affect the results of this work. In particular, in \chapt{mining} we will show that most of the detectable signal comes from the most-massive galaxies, with magnitudes well brighter than the CANDELS limiting magnitude.

\begin{table*}
%\centering
\caption{Main properties of the stacked samples.}\label{tab2}

\begin{tabular}{|r|r|r|r|r|r|r|r|r|r|}
\hline
  \multicolumn{1}{|c|}{ $z$ bin} &
        \multicolumn{1}{c|}{ $M_*/M_\odot$}  &
  \multicolumn{1}{|c|}{ N (N$^w$) } &
    \multicolumn{1}{c|}{ $\langle z^w\rangle$} &
    \multicolumn{1}{c|}{Exp.} &
    \multicolumn{1}{c|}{ $CR^{w,\rmn{TOT}}$} &
     \multicolumn{1}{c|}{ $F^{w, \rmn{TOT}}_{0.5-2\rmn{keV}}$} &  
                 \multicolumn{1}{c|}{ $SNR_{boot}$} &
             \multicolumn{1}{c|}{$N^{>CR}_{rand}$} &
             \multicolumn{1}{c|}{$SNR_{rand}$} \\      
 &&&&$10^{9}\rmn{s}$&$10^{-5}\rmn{cts\,s^{-1}}$&$10^{-16}\rmn{erg\,cm^{-2}s^{-1}}$&$\sigma$&&$\sigma$\\
  (1)&(2)&(3)&(4)&(5)&(6)&(7)&(8)&(9)&(10)\\
  \hline

$3.5\le z< 4.5$  &all                     & 1393 (1441) & 3.90 & $8.16$   & $8.17\pm4.05$ & $5.11\pm2.53$ &2.02   &$0.0091$&2.36\\
$3.5\le z< 4.5$  & $\geq1.32\times10^{9}$ & 662 (667)   & 3.91 & $3.86$   & $11.50\pm3.08$& $7.19\pm1.93$ &3.74   &$0.0000^*$&$>4.00$\\
$4.5\le z< 5.5$  & all                    & 453 (472)   & 4.90 & $2.65$   & $<2.22$       & $<1.39$       &0.87   &$0.1630$& 0.98 \\
$4.5\le z< 5.5$  &  $\geq1.86\times10^{9}$& 217 (222)   & 4.92 & $1.26$   & $3.66\pm1.79$ & $2.29\pm1.12$ &2.04   &$0.0034$&2.71  \\
$5.5\le z< 6.5$  & all                    & 230 (241)   & 5.93 & $1.35$   & $<1.61$       & $<1.01$       &$-0.34$&$0.6742$&$-0.45$ \\
$5.5\le z< 6.5$  & $\geq2.63\times10^{9}$ & 111 (113)   & 5.93 & $0.65$   & $<1.07$       & $<0.67$       &0.25   &$0.4001$&0.25  \\
  \hline

\end{tabular}\\
(1) redshift bin; (2)  mass selection: for every redshift bin we first stacked all the galaxies and then the most massive ones, defined as those with $M_*$ larger than the median in that bin; (3) number of CANDELS stacked galaxies and, in parentheses, corresponding weighted number of galaxies (Eq.~\ref{eq4}). Note that the stacked number of galaxies in the ``massive" samples is close but not equal to half the number of galaxies when no mass cut is applied. The reason is that the mass cut is the median mass of galaxies in the redshift bin, including the galaxies which are not stacked due to their proximity to (or identification with) an X-ray detected source. (4) median weighted redshift (Eq.~\ref{eq5}); (5) total exposure time; (6) total (i.e. corrected for the aperture of the extraction radii) weighted net count-rate in the soft band, (7) corresponding total flux, (8) signal-to-noise ratio (SNR) derived from the bootstrap procedure,  (9) fraction of runs of the random-position stacking returning a 
net count rate larger than that found for the sample of real galaxies, based on 10000 runs, and (10) corresponding confidence level. $^*$For this 
sample we used 50000 runs.
\end{table*}

 \subsection{Stacking results}\label{stacking}
 
\begin{figure*} 
\centering
\includegraphics[width=160mm,keepaspectratio]{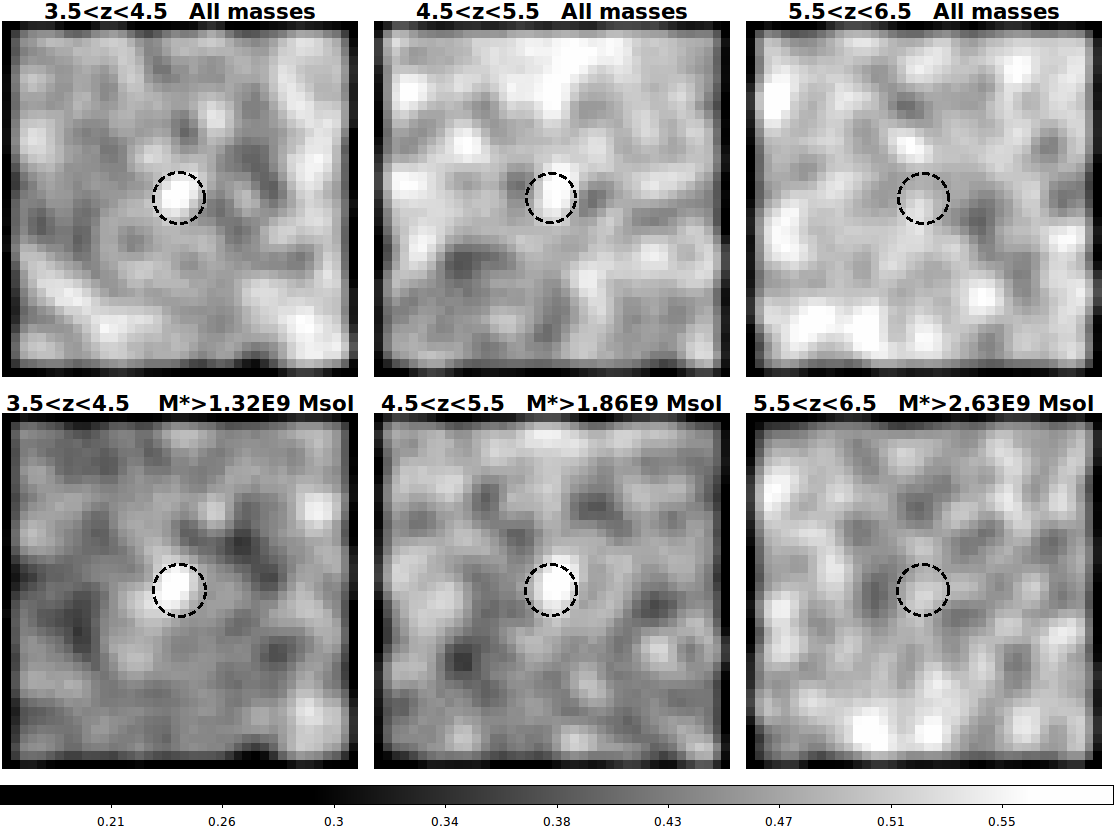}
\caption{X-ray stacked 40x40 pixels images of the samples in the observed $0.5-2\,\rmn{keV}$ band, normalized such that the pixel with the minimum (maximum) number of counts in the original image acquires a value of zero (one), and then smoothed with a Gaussian function with kernel radius of 3 pixels. The smoothing is reflected in the range of the color bar which is not $0-1$ (e.g. pixels with value $= 0$ acquire a fraction of the value of nearby pixels).  Dashed circles are centred at the position of the stacked galaxies and have radii corresponding to the median extraction radius used for sources in the corresponding redshift bin. }
\label{fig9}
\end{figure*}

  We applied the stacking procedure described in \chapt{sec1} to the 7~Ms CDF-S soft-band ($0.5-2$ keV) image for the CANDELS galaxy samples, using the \mbox{X-ray} catalogue of Luo et al. (in prep) to locate the \mbox{X-ray} detected sources. The stacked \mbox{X-ray} images in the soft band are shown in Fig.~\ref{fig9}.

\begin{figure}
\centering
\includegraphics[width=80mm,keepaspectratio]{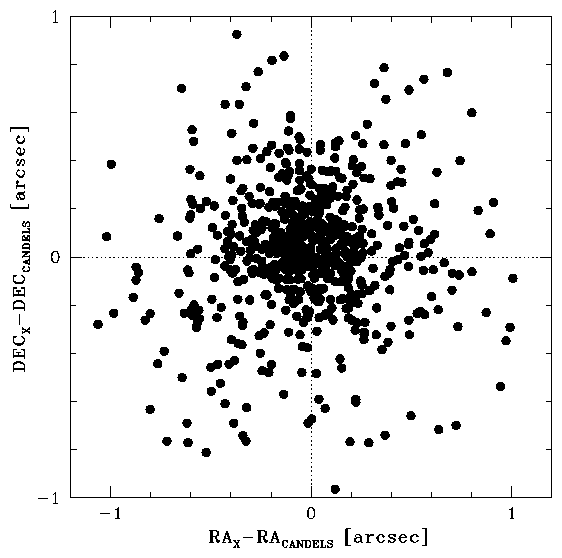}
\caption{RA and DEC offset between X-ray and CANDELS sources matched with a $1^{\prime\prime}$ radius. There is no significant systematic offset between the X-ray source positions and CANDELS galaxies.}% 

\label{offset}
\end{figure}

 We checked for a possible systematic positional offset between the 7 Ms CDF-S and CANDELS surveys by cross-matching the positions of the sources in the two catalogs within a $1''$ matching radius and computing the offset in RA and DEC between the X-ray and optical images for each matched source (Fig.~\ref{offset}). We quantified the median offset in each direction and its $1\sigma$ uncertainty through a bootstrap procedure, finding $\Delta RA_{med}=0\farcs015\pm0\farcs013$ and $\Delta DEC_{med}= 0\farcs031 \pm 0\farcs010$.
 All of these values are significantly smaller than both the radii used for the aperture photometry and the \textit{Chandra} positional uncertainty. However, while \textit{on average} there is no offset between the X-ray and optical positions, the offset distribution is scattered around zero, with a dispersion spanning from $\sim0\farcs3$ on axis to $\sim0\farcs6$ at $\theta\sim8'$. As, especially on-axis, the dispersion is comparable to the size of the region used to perform the aperture photometry, in Appendix~\ref{psf} we statistically took into account this effect in order not to underestimate the EEF.

 Tab.~\ref{tab2} reports for each stacked sample the total number of individual (N) and weighted ($\rmn{N^w}$) sources, where
 
 \begin{equation}\label{eq4}
  N^w=\sum_{i=1}^{N}\Omega_i,
 \end{equation}
the average weighted redshift

 \begin{equation}\label{eq5}
  z^w=\frac{\sum_{i=1}^{N}\Omega_i z_i}{\sum_{i=1}^{N}\Omega_i},
 \end{equation}
the total equivalent exposure time and the total (i.e. summed) weighted net count rate

 \begin{equation}\label{eq6}
  CR^w_{TOT}=\sum_{i=1}^{N}\frac{\Omega_i CR_i}{P(\theta_i)},
 \end{equation}
 where $P(\theta_i)$ is the aperture correction and takes values 0.80, 0.75, 0.60 , 0.40 at $\theta<3.5'$, $3.5'<\theta<4.25'$, $4.25'<\theta<5.5'$, $5.5'<\theta<7.8'$, respectively. The $CR^w_{TOT}$ values and their $1\sigma$ errors were derived by using a bootstrap procedure, as stated at the end of \chapt{sec1} and were then converted into total weighted fluxes.\footnote{ As we used radii corresponding to EEF=0.6, this correction is $0.6^{-1}\simeq1.67$}. 
  The conversion factor from count rate to flux was derived with PIMMS\footnote{http://heasarc.nasa.gov/docs/software/tools/pimms.html} assuming a power-law with $\Gamma=1.8$ (a reasonable value for X-ray emission related to both star-formation and AGN in the rest-frame $2-10\,\rmn{keV}$ band, up to high values of column density in the redshift ranges of interest; e.g. \citealt{Tozzi06, Mineo12}) and Galactic $N_H$. We took into account the different \textit{Chandra} Cycles during which the observations were taken by weighting the conversion factor of each pointing using the relative exposure time. Different assumptions for $\Gamma$ do not alter significantly the conversion factor and would result in fluxes well within the errors we consider.
 These flux values are reported in column 7 of Tab.~\ref{tab2}.

\begin{figure}

\includegraphics[width=85mm,keepaspectratio]{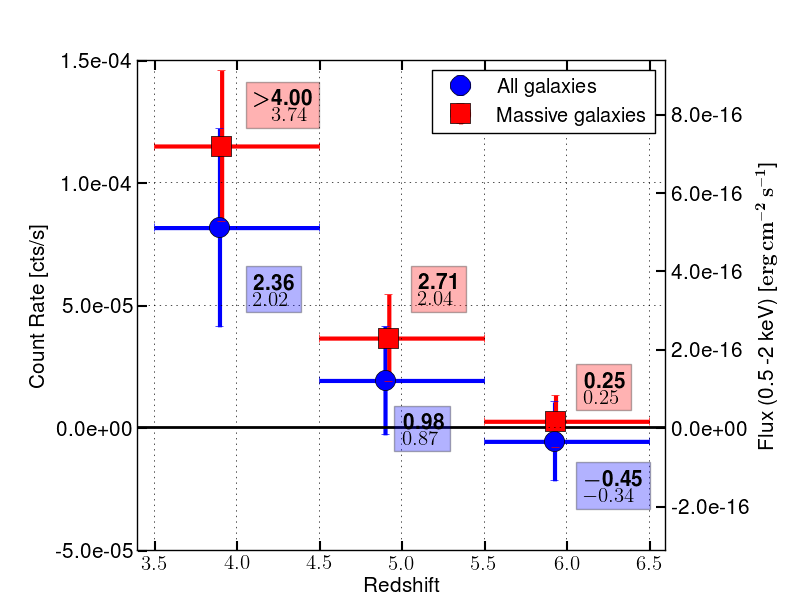}

\caption{Stacked total net count rate (left $y$-axis) and soft-band flux (right $y$-axis, derived from the count rate as described in \chapt{stacking}) with $1\sigma$ errors from bootstrapping as a function of redshift for the considered samples. Blue points include all galaxies, while red points result from stacking the high-mass galaxies (see Tab.~\ref{tab2}). The signal-to-noise ratios for each subsample computed by stacking random positions and by the bootstrap procedure are also reported in the boxes with bold and normal fonts, respectively.}% 

\label{cr}
\end{figure}

 Fig.~\ref{cr} shows the net count rates for the different samples as derived with the stacking analysis and the corresponding signal-to-noise ratios (SNR). The significance of the putative detection can be approximated by the ratio of the total net count rate and its error (column 6 in Tab.~\ref{tab2}). According to this, we detected X-ray emission with high significance ($\sim3.7\sigma$) only in the half of most-massive galaxies at $3.5\leq z<4.5$. We also report detections with lower significance ($\gtrsim2\sigma$) for the whole sample of $3.5\leq z<4.5$ galaxies and for the half of most-massive galaxies at $4.5\leq z<5.5$.

\begin{figure}

\includegraphics[width=80mm,keepaspectratio]{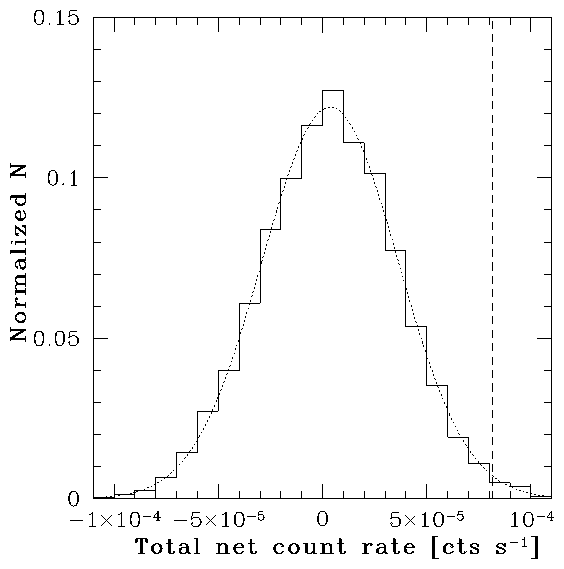}

\caption{Normalized distribution of the total net count rate derived by running 10000 times the stacking procedure at the position of the $3.5\leq z<4.5$ sample, perturbed by a random value up to $20''$ in each direction. The dotted curve is the best-fitting Gaussian function. The vertical dashed line marks the total net counts of the real sample of galaxies in the $3.5\leq z <4.5$ bin.}% 

\label{net}
\end{figure}

  A more accurate measurement of the significance level of the (non) detections, following the definition itself of ``significance level", can be obtained by repeatedly stacking $N$ random positions in the field, with $N$ equal to the number of sources in a sample, and counting how many times the derived count rate is larger than the one obtained for the real sources \citep[e.g.][]{Brandt01}. Therefore, for each sample we ran the stacking procedure 10000 times, perturbing the coordinates of the galaxies by a random value up to $20''$ in each direction. Doing this we efficiently create a list of random positions to be stacked which, at the same time, statistically maintain the spatial distribution of the real sources. For every run we computed the total net count rate in the same way as for the real sources (i.e. through the bootstrap procedure). We report in Fig.~\ref{net} the distribution of 
the 10000 total net count rates obtained for the $3.5\leq z<4.5$ sample as an example. The distribution is centred very close to zero, which is evidence that the procedure is not affected by any apparent bias (see \citealt{Cowie12}, \citealt{Treister13}). The number of runs which gives a total net count rate smaller than the ones derived for the samples of real galaxies is an independent estimate of the significance of a detection. We report these numbers for each sample in the {\bf columns 8 and 10} of Tab.~\ref{tab2} and in Fig.~\ref{cr}. According to this procedure, the significance of the detection of massive galaxies at $z\sim4$ increases to $>4\sigma$. The detection significance for the whole $z\sim4$ sample increases as well ($2.4\sigma$). We also report the detection with the highest significance ($2.7\sigma$, corresponding to a $99.7\%$ confidence level) of stacked X-ray emission from massive ($M_*>1.86\times10^9\rmn{M_\odot}$) galaxies at $z\sim5$. This is indeed the detection with the highest 
significance of X-ray emission from galaxies at such a high redshift. Previously, \cite{Cowie12} reported a tentative ($\sim1\sigma$) detection of X-ray emission from stacked galaxies at $z\sim5$ in the CDF-N, although the significance decreased once they included the {\mbox 4 Ms CDF-S} data. We attribute this 
result to the increased exposure 
of the \chandra\, observations coupled with the new deep optical data available in this field. 
 In Appendices~\ref{PDF} and \ref{LBG} we investigate the impact of the photometric-redshift uncertainties and the particular selection approach we used on our results, finding that they appear robust against these two issues. In particular, in Appendix \ref{LBG} we will apply our stacking analysis to a sample of Lyman-Break Galaxies (LBG) at $z=4-8$. Considering the sample sizes and the use of the deepest X-ray data available, those are the best constraints to date on the stacked X-ray emission from galaxies at such redshifts.


\subsection{The dominant undetected X-ray population}\label{mining}
\begin{figure} 
% \centering
\includegraphics[width=80mm,keepaspectratio]{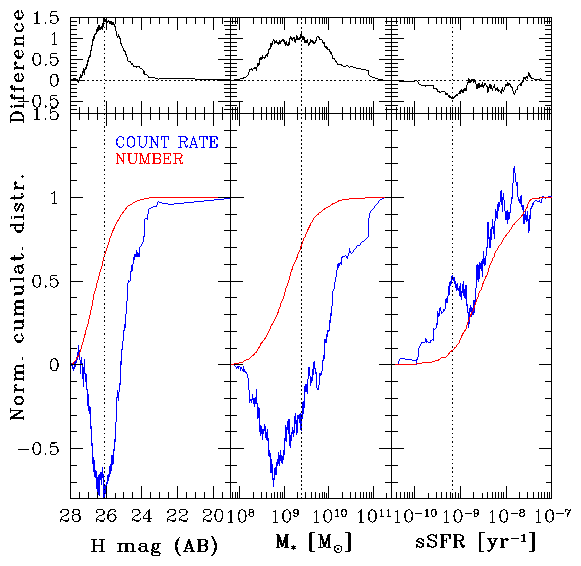}
\caption{Normalized cumulative distribution of the number of stacked galaxies (red curves) and net count rate (blue curves) as a function of \Hband magnitude (left panel), stellar mass (central panel) and sSFR (right panel) for the $3.5\leq z < 4.5$ bin. The upper inserts show the difference between the two curves (solid line) and the zero level (horizontal dotted line). The vertical dotted line marks the location of the maximum difference.}
\label{subsamples}
\end{figure}

In the previous sections we found that most, if not all, of the signal from galaxies at $3.5\leq z<4.5$ and $4.5\leq z<5.5$ comes from the most massive half of the samples. However, the chosen mass threshold (i.e. the median mass of the sample) could be not the one that would return the highest SNR. Moreover, we cannot be sure that mass is the primary physical parameter driving the \mbox{X-ray} emission. To address these issues, in Fig.~\ref{subsamples} we plotted the normalized cumulative distribution of the number of stacked sources and the stacked signal as a function of \Hband magnitude, stellar mass and sSFR for the $3.5\leq z<4.5$ sample. The sSFR was preferred over the SFR since it is a quantity defined per unit stellar mass. We could derive the distribution in count rate as we know the contribution of the individual stacked galaxies (which can also be negative) to the total signal from our aperture photometry procedure. For 
instance, the left panel of Fig.~\ref{subsamples} shows that $\sim90\%$ of the stacked galaxies at $3.5\leq z<4.5$ have $H>25$, but their cumulative contribution to the final net signal is null. In principle, if every galaxy contributed to the stacked signal in the same way, with no dependence on magnitude, mass or sSFR, the blue and red curves would overlap, aside from the statistical noise on the measured count rate. The different behaviour of the two curves in two out of the three cases tells us that the stacked X-ray emission is actually dominated by luminous and massive galaxies (and is not affected by preferential sSFR values). 

The difference between the cumulative distribution of the number of galaxies and the relative count rate includes information about how much the stacked emission depends on magnitude, mass, and sSFR. These difference values are plotted in the upper inserts of Fig.~\ref{subsamples}. In particular, the location of the maximum difference (dotted lines in Fig.~\ref{subsamples}) is a good estimate of the cut in magnitude, mass, or sSFR to be applied in order to get the maximum signal from the minimum number of galaxies. For instance, in the second panel, the maximum difference ($\sim1$) occurs at $M_*\sim2.4\times10^{9}M_\odot$, corresponding to $\sim70\%$ of the stacked galaxies and a cumulative negative flux, which means that stacking galaxies with $M_*>2.4\times10^{9}M_\odot$, $\approx30\%$ of the sample, would return a stronger signal than stacking the whole sample ($\sim140\%$), as less-massive galaxies preferentially returned negative count rates. The maximum difference is lower when investigated as 
a function of sSFR than for the magnitude and mass cases. This means that the sSFR is not the main driver of the \mbox{X-ray} emission. The maximum difference is very similar for the \Hband magnitude and mass cases; this is not unexpected as the stellar mass from SED fitting is typically determined from the optical/near-IR part of the spectrum.

\begin{figure} 
\centering
\includegraphics[width=80mm,keepaspectratio]{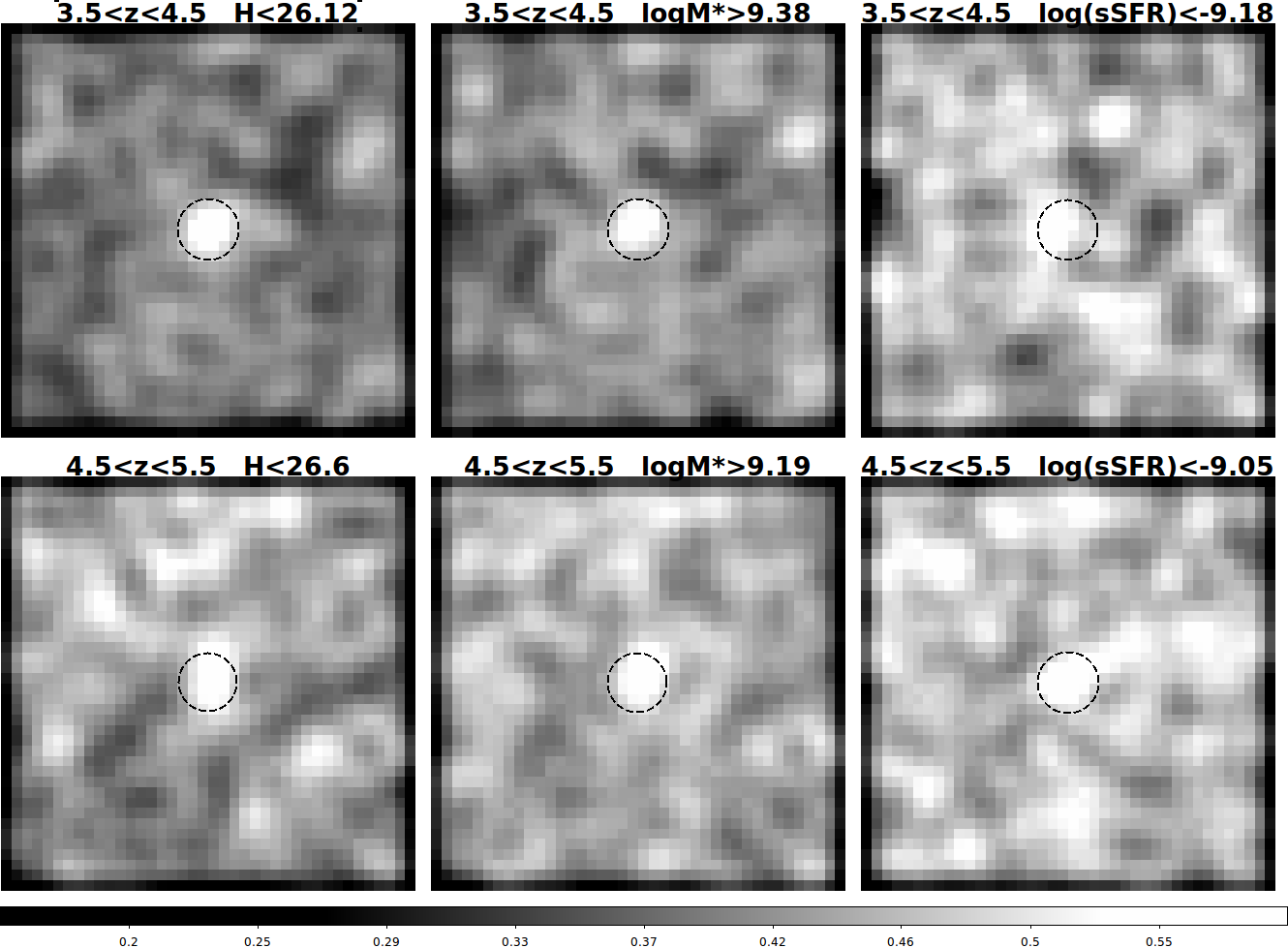}
\caption{ X-ray stacked 40x40 pixel images of the samples selected applying the magnitude, mass, and sSFR cuts (from left to right) in the $3.5\leq z<4.5$ (upper row) and $4.5\leq z<5.5$ (lower row) derived in \chapt{mining} in the observed $0.5-2\,\rmn{keV}$ band. Images are normalized as in Fig.~\ref{fig9} as described in \chapt{stacking} . A Gaussian smoothing with a kernel radius of 3 pixels was applied. The scale is linear. Dashed circles are centred at the position of the stacked galaxies and have radii corresponding to the median extraction radius used for sources in the corresponding redshift bin.}
\label{subsamples_stack}
\end{figure}

 We used the magnitude, mass, and sSFR values corresponding to the maximum difference between the number of galaxies and the net count rate distributions as cuts to identify the galaxies in our $3.5\leq z<4.5$ and $4.5\leq z<5.5$ samples that contribute most to the stacked signal. We then repeated the stacking procedure for these subsamples. We note that the maximum difference for the sSFR case is negative, which means that galaxies with low sSFR contribute slightly more than those with high sSFR to the total signal. In this case, we therefore stacked galaxies with sSFR lower than that corresponding to the maximum difference. Fig.~\ref{subsamples_stack} shows the stacked images in the soft-band of the subsamples. Tab.~\ref{max_sig} reports the cuts in magnitude, stellar mass and sSFR used for galaxies at $3.5\leq z<4.5$ and $4.5\leq z<5.5$ applied to derive the maximum SNR, the fraction of stacked galaxies with respect the parent sample, the relative net-count rate and the SNR computed by the bootstrap 
procedure. For the $4.5\leq z<5.5$ bin, we also computed the SNR by stacking $N$ random positions, with $N$ equal to the number of stacked sources in each subsamples, and counting the number of stacking run which returned a lower signal than the measured one, as we did in \chapt{stacking}. These values are reported in Tab.~\ref{max_sig}. In particular, applying the selection method described in this section, the significance of the signal from bright ($H<26.6$) and/or massive ($\rmn{log}M_*>9.19$) galaxies at $z\sim5$ increases to $>3\sigma$.

However, these detection significances may be biased upwards because the subsamples were constructed to maximize SNR using prior information on count rates. To independently estimate the significance, we used a set of 1000 4-fold cross-validation tests \citep{Efron93}, in a similar way as \cite{Xue12}: first we 
randomly split the full 
sample into 4 approximately equal-sized subsamples; we then excluded one subsample and repeated the procedure above jointly on the 3 remaining subsamples in order to derive the magnitude (or mass or SFR) cut, which was then applied to the excluded subsample to select $\sim1/4$ of galaxies to be stacked, i.e. the threshold in the parameter of interest was evaluated using $\sim3/4$ of the sample and then applied to the remaining subsample. The same was repeated for the other 3 subsamples, and the final sample (selected using different thresholds in the parameter of interest) was stacked and a SNR was derived. The procedure was repeated 1000 times, allowing us to derive a distribution of SNR, whose median is the actual SNR of the detection. As noted by \cite{Xue12}, these SNRs are likely 
underestimated, as we only used $\sim3/4$ of the sample to determine each selection. The real SNR therefore lies in the ranges defined by the SNR computed through the bootstrap procedure and the 4-foldcross-validation tests. The resulting SNR are reported in Tab.~\ref{max_sig}. These numbers indicate that the X-ray signal is mainly driven by the mass/magnitude of the galaxies.

The average stacked X-ray fluxes we obtained for the whole samples in \chapt{stacking} (obtained by dividing column 7 by column 3 of Tab.~\ref{tab2}) are \mbox{$\gtrsim3\times10^{-19}\rmn{erg\,cm^{-2}s^{-1}}$}, similar to the fluxes of sources expected to be individually detected by \textit{X-Ray Surveyor}  (see \chapt{conclusions}), which will have an expected flux limit for a 4 Ms exposure of $\sim3\times10^{-19}\rmn{erg\,cm^{-2}s^{-1}}$ \citep[see sec. 6 of ][]{Weisskopf15}. However, in this section we have shown that the detected signal is due to a small fraction of optically-bright, massive galaxies. For instance, as described in the first paragraph of this section, the detected signal at $z\sim4$ is entirely due to the $\approx10\%$ of most massive galaxies, while the remaining $\approx90\%$ of the sample has a null cumulative stacked emission. Therefore, we are effectively sampling a population of galaxies with an average X-ray flux a factor of $\sim10$ larger than the \textit{X-Ray Surveyor} flux 
limit for individually detected sources, thus already placing useful
constraints on the scientific outcomes expected by \textit{X-Ray Surveyor}.

\begin{table}
%\centering
\caption{Stacking results for the magnitude, stellar mass and sSFR subsamples at  $3.5\leq z < 4.5$ and $4.5\leq z < 5.5$ defined by the reported thresholds, derived as described in \chapt{mining}. We also report the fraction of sources in these subsamples with respect to the parent sample. The given ranges for SNR values are derived through a 4-fold cross-validation test and from the bootstrap procedure. For the $4.5\leq z < 5.5$ bin we also report the SNR (in parentheses) derived from stacking random positions in the field.  }\label{max_sig}
\begin{tabular}{rrr}

  \multicolumn{1}{c}{{}} &
 \multicolumn{1}{c}{{\bf $3.5\leq z < 4.5$ }} &
 \multicolumn{1}{c}{{\bf $4.5\leq z < 5.5$ }}\\ 
  \hline
  $H<$& 26.12 & 26.60\\
 Fraction& 0.36 & 0.51\\
 CR ($10^{-4}\mathrm{cts\,s^{-1}}$) & $1.42\pm 0.27$&$0.42\pm 0.18$\\
SNR & 4.92--5.27&2.36--2.38 (3.24)\\
\hline
  log$M_*>$& 9.38 & 9.19\\
 Fraction& 0.29 & 0.55\\
 CR ($10^{-4}\mathrm{cts\,s^{-1}}$) & $1.13\pm 0.25$&$(0.42\pm 0.19)$\\
SNR & 3.86--4.55&2.10--2.23 (3.16)\\
\hline
  log(sSFR)$<$& $-9.18$ & $-9.05$\\
 Fraction& 0.09 & 0.18\\
 CR ($10^{-5}\mathrm{cts\,s^{-1}}$) & $4.70\pm 1.57$&$(1.42\pm 0.27)$\\
SNR & 1.85--2.99&0.93--2.03 (2.13)\\

\end{tabular}\\
\end{table}

\section{Nuclear accretion and star formation properties of high-redshift galaxies}\label{results}
In this section we use the stacking results from individually X-ray-undetected galaxies to infer information about the BHAD, SFRD and faint end of the XLF at $3.5\leq z<6.5$, under different assumptions about the relative contributions of the nuclear and stellar X-ray emission. We also include separately results from a sample of X-ray detected, high-redshift AGN.

 \subsection{$\rmn{L_X-SFR}$ relations}\label{Lx_sfr}
 X-ray emission from galaxies is due to a combination of nuclear accretion and stellar processes, primarly in the form of \mbox{X-ray} binaries (XRBs). In particular, emission from high-mass XRB (HMXB) and low-mass XRB (LMXB) is related to the SFR and $M_*$ of the host galaxy, respectively \citep[e.g.][]{Lehmer10,Mineo12, Zhang12}. At the lowest fluxes probed by the 4~Ms CDF-S, \cite{Lehmer12} found that about half of the X-ray detected sources are star-forming galaxies, whose relative number is expected to increase further at even lower fluxes (such as those probed by this work), overtaking the number of AGN. It is therefore necessary to consider the XRB contribution in our investigation of the X-ray emission from galaxies. However, the relation between SFR and X-ray emission is not observationally constrained in the redshift range probed by this work. 
 
 We therefore use two different $L_X-SFR$ relations: a simple local-universe calibration \citep{Ranalli03} with no evolution, and models dependent on galaxy properties with redshift evolution \citep{Fragos13}.  The \cite{Fragos13} models are constructed by applying the Startrack XRB-population synthesis code \citep{Belczynski02, Belczynski08} to the Millenium II cosmological simulation \citep{Guo11}, and predict the $L_X\rmn{(HMXB)}-SFR$ and $L_X\rmn{(LMXB)}-M_*$ relations as a function of redshift up to $z\sim20$, calibrated in such a way to match the observed relations in the local universe. We will use the \cite{Fragos13} theoretical relation corresponding to their model 269, which has been found by Lehmer et al. (accepted) to be the one that best agrees with observational data up to $z\sim2.5$ and the slight redshift evolution 
observed by \cite{Basu-Zych13_2}, at the same time providing a physically motivated extrapolation to $z=3.5-6.5$. For instance, at high redshift the typical time-scale needed to form LMXB after a burst of star formation is comparable to the age of the Universe. Moreover, the number of HMXB and their mass distribution depend on metallicity, which evolves with redshift. These considerations imply a flattening of the $L_X\rmn{(LMXB)}-M_*$ relation at $z>3$, followed by a decline at $z>5-6$, and a mild positive evolution of the $L_X\rmn{(HMXB)}-SFR$ relation with increasing redshift. While these behaviours are predicted by the \cite{Fragos13} theoretical modelling (see their fig. 5), the $L_X\rmn{(HMXB)}-SFR$  and $L_X\rmn{(LMXB)}-M_*$ relations have not been constrained observationally at $z\gtrsim3$ (e.g. Lehmer et al. accepted).
 We corrected the \cite{Ranalli03} and \cite{Fragos13} relations in order to match the IMF assumed by \cite{Santini15}.

  \subsection{ Assumptions on the relative contribution from AGN and SF}\label{relative_AGN_SF}
  In this section, we describe how we account for the contribution from both AGN and X-ray binaries to the stacked X-ray emission. As we do not know the relative strength of these two classes of processes, we consider 5 cases:
 \begin{enumerate}
 \item All the X-ray emission is due to AGN activity. 
 \item All the X-ray emission is due to processes related to star formation (SF), which scales with $L_X$ following the \cite{Ranalli03} relation.
 \item Same as (ii), but using the \cite{Fragos13} relation. In this case we assume the mean $M_*$ of each sample.
  \item The X-ray emission is due to a mixture of AGN and SF, which scales with $L_X$ following the \cite{Ranalli03} relation.
  \item Same as (iv), but using the \cite{Fragos13} relation.

 \end{enumerate}
 Case i gives us an upper limit on AGN emission from high-redshift X-ray-undetected galaxies. Similarly, cases ii and iii give upper limits on the SFR. Case iv and v represent the more realistic cases in which AGN and XRB contribute simultaneously to the total X-ray emission. 
 For the last two cases we assume the mean SFR and $M_*$ of the considered sample to obtain the X-ray emission due to XRB. Then we subtracted it from the total X-ray flux derived from the stacking analysis to estimate the emission due to nuclear accretion.

 \subsection{Correcting for obscuration}\label{obsc_corr}

 In \cite{Vito13,Vito14} we showed that at $z\approx3.5$ approximately half of the X-ray selected AGN are obscured by a column density $\rmn{log}N_H>23$ with a median value of $\rmn{log}N_H\approx23.5$ and no strong dependence on luminosity. When dealing with AGN, we thus assume that half of the sample is obscured by $\rmn{log}N_H\approx23.5$ and that all sources have identical intrinsic (i.e. unabsorbed) flux. Under these assumptions, the intrinsic total soft band flux is:
 \begin{equation}\label{abs}
  F_{0.5-2\,\rmn{keV}}=\frac{2\times  F_{0.5-2\,\rmn{keV}}^{obs}}{(1+c(z))},
 \end{equation}

 \noindent where $c$ is the transmission factor due to absorption by a column density $\rmn{log}N_H\approx23.5$. Assuming an intrinsic slope of $\Gamma=1.8$, $c(z)$ ranges from $\approx0.3$ at $z=4$ to $\approx0.5$ at $z=6$. We always apply Eq.~\ref{abs} to the observed flux ascribed to AGN activity, thus correcting it by a factor $1.3-1.5$.

 \subsection{X-ray detected high-redshift AGN}\label{Xdet}
 In order to derive a more comprehensive picture of the accretion properties of high-redshift galaxies, we also considered sources detected in the 7 Ms CDF-S (Luo et al. in prep.) that fall within the same sky field considered so far (i.e. CANDELS/GOODS-S restricted to $\theta<7.8'$ from the \mbox{CDF-S} average aim point). 
 We used the spectroscopic redshifts provided by \cite{Vito13} and Luo et al. (in prep.) and photometric redshifts from \cite{Skelton14} and \cite{Hsu14}.\footnote{In the updated version retrievable at http://www.mpe.mpg.de/XraySurveys/CDF-S/CDF-S\_X\_photoz.html} In 
 particular, \cite{Hsu14} used hybrid (galaxy plus AGN) templates for X-ray detected sources in the 4~Ms CDF-S. For those sources the \cite{Hsu14} redshifts are therefore expected to be more reliable than the \cite{Santini15} ones. In total, there are $11$ AGN at $3.5\leq z<6.5$ detected in the reduced \mbox{CANDELS/GOODS-S} field in the 7 Ms CDF-S.

 In our field, 10 sources out of the $691$ X-ray detected objects ($\sim1\%$) have no optical counterpart and thus no redshift information. 
  Including a correction for redshift incompleteness to the results derived for individually-detected sources is beyond the scope of this paper, which is focused on the population of individually-undetected galaxies. However, as will be shown in \chapt{BHAD} and \chapt{discussion_bhad}, such a correction would only strengthen the results.

 X-ray spectra were extracted using the ACIS Extract software\footnote{http://www2.astro.psu.edu/xray/docs/TARA/ae\_users\_guide.html} and fitted with XSPEC.\footnote{https://heasarc.gsfc.nasa.gov/xanadu/xspec/} As in \cite{Vito13}, we used a simple spectral model in order to derive basic spectral parameters such as column density, flux and intrinsic luminosity. In particular, we assumed an absorbed power-law with intrinsic photon index fixed to $\Gamma=1.8$, including Galactic absorption \citep{Kalberla05}. Fitting  results are reported in Tab.~\ref{tab_Xdet}. The use of more complex models is precluded for most of the considered sources because of the limited photon counting statistics. Moreover, our aim is to derive reasonable basic estimates of the 
intrinsic luminosity rather than to perform a detailed spectral analysis. A dedicated analysis of the population of X-ray detected, high-redshift AGN will be the topic of a future work.

\begin{table*}
\scriptsize
\caption{Main properties and best-fitting spectral parameters of the X-ray detected $3.5\leq z<6.5$ AGN in the CANDELS field within $\theta<7.8'$ of the CDF-S average aim point}\label{tab_Xdet}
\begin{tabular}{cccccc}
\hline
  \multicolumn{1}{ c }{CXOCDFS J} &  
  \multicolumn{1}{ c }{$z$} &
  \multicolumn{1}{ c }{$z_{ref}$} &
  \multicolumn{1}{c }{$N_H$} &
  \multicolumn{1}{c}{$F_{0.5-2} $} &
  \multicolumn{1}{c}{$L_{2-10} $}  \\
   &&&$10^{22}\rmn{cm^{-2}}$&$\rmn{erg\,cm^{-2}\,s^{-1}}$&$\rmn{erg\,s^{-1}}$\\
  (1) &(2) &(3) &(4) &(5) &(6)  \\
\hline                                                                                    %Kbol    %Lbol
033207.3$-$274942 & $3.515_{-0.045}^{+0.045}$  &  P1      & $13^{+7}_{-6}$       & $5.94\times10^{-17}$ & $1.68^{+0.61}_{-0.47}\times10^{43}$ \\ %XID 165
033209.8$-$275015 & $4.651_{-1.861}^{+1.199}$   & P1       &  $49^{+13}_{-12}$    &$7.80\times10^{-17}$  &$6.77^{+1.09}_{-1.01}\times10^{43}$ \\ %XID 195
033211.3$-$275213 & 3.740   &  S1      & $30^{+2}_{-2}$       & $1.36\times10^{-15}$ & $5.57^{+0.17}_{-0.16}\times10^{44}$ \\  %XID 214
033213.0$-$274351 & $5.840_{-0.156}^{+0.107}$ &  P2      & $44^{+39}_{-37}$       & $3.84\times10^{-17}$ & $3.69^{+1.49}_{-1.32}\times10^{43}$ \\  %XID 238
033216.8$-$275043 & 3.712   &  S2      & $13^{+20}_{-10}$     & $2.39\times10^{-17}$ & $7.16^{+3.31}_{-2.10}\times10^{42}$ \\ %XID 299
033218.8$-$275135 & 3.660   &  S1      & $101^{+10}_{-9}$     & $8.14\times10^{-17}$ & $1.59^{+0.15}_{-0.13}\times10^{44}$ \\ %XID 337
033223.1$-$275029 & $3.501_{-0.071}^{+0.079}$   &  P1      & $124^*$     & $<10^{-17}$ & $<7.61\times10^{42}$ \\ %XID 414
033225.8$-$275507 & $4.682_{-0.224}^{+0.192}$    &  P2      & $13^{+7}_{-6}$   & $1.96\times10^{-16}$ & $8.27^{+1.15}_{-1.02}\times10^{43}$ \\ %XID 464
033229.8$-$275105 & 3.700   &  S1      & $81^{+5}_{-5}$       & $1.94\times10^{-16}$ & $2.64^{+0.15}_{-0.15}\times10^{44}$ \\ %XID551
033233.4$-$275227 & $3.876_{-0.376}^{+0.014} $  &  P1      & $39^{+15}_{-13}$     & $4.09\times10^{-17}$ & $2.53^{+0.63}_{-0.52}\times10^{43}$ \\ %XID623
033241.8$-$275202 & 3.610   &  S1      & $7^{+1}_{-1}$        & $1.55\times10^{-15}$ & $3.48^{+0.18}_{-0.17}\times10^{44}$ \\ %XID774

\hline
 \hline
\end{tabular}\\
\normalsize
(1) Official \textit{Chandra} name of the X-ray source; (2) redshift; (3) redshift reference - S1, S2: spectroscopic redshift from Luo et al. (in prep.) and \cite{Vito13}, respectively; P1, P2 : photometric redshift from \citet[http://www.mpe.mpg.de/XraySurveys/CDF-S/CDF-S\_X\_photoz.html]{Hsu14} and \citet{Skelton14}, respectively; (4) best-fitting column density. $^*$ The low photon-counting statistics of this object prevented us from leaving the column-density parameter free, and thus we fixed it to the value reported in the Luo et al. (in prep.) catalog. (5) soft-band flux; (6) hard-band ($2-7\,\rmn{keV}$), intrinsic (i.e. rest-frame and corrected for absorption) luminosity. All errors are at the $1\sigma$ confidence level. Errors on luminosity account only for the error on the power-law normalization.  

\end{table*}

 \subsection{Black Hole Accretion Rate Density (BHAD)}\label{BHAD}
 
 The BHAD is defined as 
 
 \begin{equation}\label{psi1}
  \Psi_{bhar}(z)= \int \frac{(1-\varepsilon)}{(\varepsilon c^2)} L_{bol,AGN} \phi(L_{bol,AGN}, z)\rmn{dlog}L_{bol,AGN}
 \end{equation}

\noindent where $L_{bol,AGN} = K_{bol}(L_{X,AGN})L_{X,AGN}$ is the bolometric luminosity, $K_{bol}$ is the bolometric correction suitable for the hard X-ray band ($2-7\,\rmn{keV}$), $\varepsilon$ is the radiative efficiency and $\phi(L_{bol,AGN}, z)$ is the AGN bolometric luminosity function We assumed $\varepsilon=0.1$ and $K_{bol}=10$ (which is appropriate for $\rmn{logL_X}\lesssim42$ AGN according to \citealt{Marconi04} and \citealt{Lusso12}) with no dependence on X-ray luminosity for the stacked samples. Under these assumptions, Eq.~\ref{psi1} becomes
 
 \begin{multline}\label{psi2}
    \Psi_{bhar}(z)=   \\  
    \frac{(1-\varepsilon)K_{bol}}{(\varepsilon c^2)}\int  L_{X,AGN} \phi(L_{X,AGN},z)\rmn{dlog}L_{X,AGN} \, \rmn{[cgs]}
 \end{multline}

 \noindent The integral is the co-moving AGN emissivity per unit volume ($U_{AGN}$):
 
 \begin{equation}
  U_{AGN}\simeq\frac{\rmn{L^{TOT}_{AGN}}}{V_c^{'CANDELS}}
 \end{equation}

 \noindent where $V_c^{'CANDELS}$ is the comoving volume covered by the CANDELS field in the redshift range of interest, after correcting it for the area lost by considering only $\theta<7.8'$ and having masked regions where X-ray detected sources are located ($\sim0.004\,\rmn{deg^2}$, corresponding to an area loss of $\sim10\%$). We computed $\rmn{L^{TOT}_{AGN}}$ (the luminosity due to nuclear accretion in the stacked galaxies) from the fluxes reported in Tab.~\ref{tab2}, under the assumptions described in \chapt{relative_AGN_SF} and taking into account the correction for obscuration (\chapt{obsc_corr}), at the mean redshift of each sample.

\begin{figure*} 
\centering
\includegraphics[width=160mm,keepaspectratio]{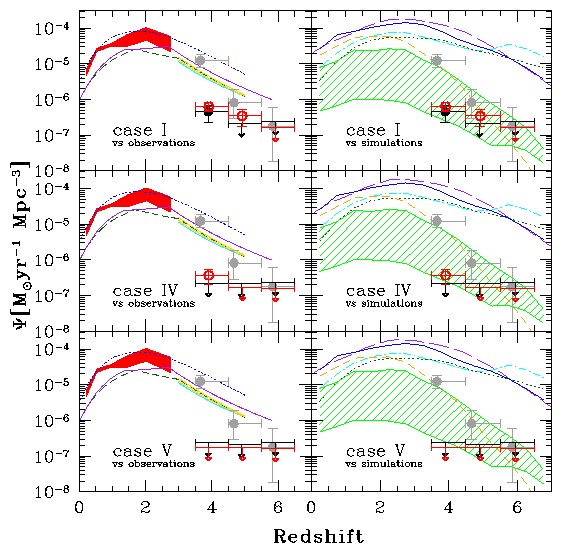}

\caption{ {\bf Constraints on the BHAD} as a function of redshift derived from the stacking analysis in cases i (upper row), iv (middle row) and v (lower row) for all galaxies (black points and upper limits) and for the most massive ones only (red points and upper limits). Grey points show the contribution of the X-ray detected AGN. Their error bars account for both the uncertainties on the X-ray luminosities (Tab~\ref{tab_Xdet}) and the statistical errors on the number of detected sources. No correction for redshift incompleteness was applied. Our findings are compared with observational results (left panels) for AGN samples (thus directly comparable with the grey points) from \citet[red shaded region]{Delvecchio14}, \citet[dashed green line]{Ueda14}, \citet[yellow-striped region]{Vito14}, \citet[dotted blue line]{Aird15}, \citet[solid cyan line]{Georgakakis15} and \citet[solid purple line]{Miyaji15} and with results from simulations (right panels) of \citet[dot-dashed cyan line]{Lodato06}, \citet[dotted 
green line]{Volonteri10}, \citet[short-dashed orange line, representing light seeds, and green-striped region, encompassing the results for 
massive seeds under different assumptions about accretion]{Bonoli14}, \citet[long-dashed purple line]{Sijacki15} and \citet[solid blue line]{Volonteri16}. See \chapt{BHAD} for details.} 

\label{fig11}
\end{figure*}

 The BHAD values for each redshift bin and considered case are reported in Tab.~\ref{tab3}. Fig.~\ref{fig11} shows the results as a function of redshift for the three cases of \chapt{relative_AGN_SF} in which we considered an AGN contribution to the X-ray emission, compared with some previous observational results in the literature (panels in the left column): \cite{Delvecchio14} derived the BHAD for a sample of \textit{Herschel}-detected AGN using a SED-fitting procedure, while \cite{Ueda14}, \cite{Vito14}, \cite{Aird15}, \cite{Georgakakis15} and \cite{Miyaji15} derived the X-ray luminosity function for AGN. We converted those luminosity functions into BHADs using Eq.~\ref{psi1}. In this process we had to assume a $K_{bol}$. \cite{Lusso12} derived the corrections for converting from the $2-10\,\rmn{keV}$ to the bolometric luminosity due to accretion only, while \cite{Hopkins07} derived corrections which refer to the total, observed 
bolometric luminosity of AGN hosts. As we are 
interested in 
emission due only to 
accretion, we use the \cite{Lusso12} $K_{bol}$. However, we integrated the X-ray luminosity functions up to luminosities exceeding the luminosity range probed by that work ($\rmn{log(L_{bol}/L_\odot)}\sim10-13$). Therefore, at larger luminosities we used the \cite{Hopkins07} corrections, also noting that, at QSO-like luminosities, $\rmn{L_{bol}}$ of the host galaxies is dominated by the accretion component. The difference between the BHAD from \cite{Aird15} and the BHADs derived from the other X-ray based studies is largely due to the different $K_{bol}$ used, as \cite{Aird15} assumed the \cite{Hopkins07} conventions at all luminosities. 
The panels in the right column of Fig.~\ref{fig11} show the comparison between our findings and results based on simulations which assume different flavours of SMBH seeding and growth mechanisms. In particular, massive seeds ($M_{BH}\sim10^5\,\rmn{M_\odot}$) are used by \cite{Lodato06}, \citet[orange line]{Bonoli14}, \cite{Sijacki15} and \cite{Volonteri16}, while the BHAD of \cite{Volonteri10} and \citet[yellow region]{Bonoli14} are due to accretion onto light seeds (Pop III remnants, $M_{BH}\sim10^2\,\rmn{M_\odot}$). These BHAD curves have been derived differentiating the BH mass-density curves presented by all of these works (except for \citealt{Sijacki15}, who directly report the BHAD). While these curves should be compared to the total BHAD, this is consistent (i.e. within the errors) with the BHAD due to AGN alone. In fact, the contribution of the stacked galaxies is more than one order of magnitude lower at $z\sim4$ and is at most equal at higher redshifts than the BHAD due to individually X-ray 
detected sources.
 
   We also show the results from the sample of X-ray detected AGN in CANDELS/GOODS-S described in \chapt{Xdet}, but we note that they should be treated with caution, as they are easily affected by small-number statistics and cosmic variance. This is why the BHAD for detected AGN is kept separate from the BHAD derived from the stacked signal, which should be less sensitive to those issues. 
Also, we did not account for the stellar contribution to the \mbox{X-ray} luminosity of the individually detected sources, as they are expected to be strongly dominated by the AGN, their \mbox{X-ray} luminosity being well above the maximum luminosity produced by X-ray binaries ($\rmn{log}L_X\sim42$, e.g. \citealt{Ranalli03, Ranalli05}).
 For these X-ray detected sources, we use the appropriate $K_{bol}$ corresponding to the observed luminosity and applied Eq.~\ref{psi2}. The implications of the results shown in Fig.~\ref{fig11} are discussed in \chapt{discussion_bhad}.

\begin{table*}
%\centering
\caption{Black hole accretion rate density in units of $10^{-7}\rmn{M_\odot yr^{-1} Mpc^{-3}}$ for the X-ray detected AGN and for each stacked sample. }\label{tab3}
\begin{tabular}{|r|r|r|r|r|r|r|r|}
\hline
  \multicolumn{1}{|c|}{{\bf bin}} &
    \multicolumn{1}{|c|}{{\bf $M_*/M_\odot$}} &
  \multicolumn{2}{|r|}{{\bf BHAD }} \\ 
  \cline{3-6}
   &&X-ray detected &\multicolumn{3}{c}{ X-ray undetected} \\
   \cline{4-6}
  && &case i & case iv & case v &\\

  \hline
$3.5\le z< 4.5$ & all                  & $123.50^{+61.00}_{-43.00}$& $4.50\pm2.23$ & $<2.23$       & $<2.35$\\
$3.5\le z< 4.5$ & $\geq1.32\times10^9$ &                           & $6.37\pm1.69$ & $3.69\pm1.69$ & $<1.76$\\
$4.5\le z< 5.5$ & all                  & $8.16^{+10.80}_{-5.29}$   & $<2.01$       & $<2.12$       & $<2.16$\\
$4.5\le z< 5.5$ & $\geq1.86\times10^9$ &                           &$3.47\pm1.68$  & $<1.73$       & $<1.77$\\
$5.5\le z< 6.5$ & all                  & $1.79^{+4.18}_{-1.61}$    & $<2.39$       & $<2.41$       & $<2.42$\\
$5.5\le z< 6.5$ & $\geq2.63\times10^9$ &                           & $<1.60$       & $<1.61$       & $<1.63$\\

  \hline

\end{tabular}\\
\end{table*}

 \subsection{Star Formation Rate Density (SFRD)}\label{SFRD}
 
 We define the SFRD as:
 \begin{equation}\label{SFRD_eq}
  \rho_{SFR}=\frac{SFR^{TOT}}{V_c^{'CANDELS}}=\frac{\rmn{N}^w\langle SFR\rangle}{V_c^{'CANDELS}} \,\rmn{M_\odot yr^{-1}Mpc^{-3}}.
 \end{equation}
 
\begin{figure} 
\centering

\includegraphics[width=90mm,keepaspectratio]{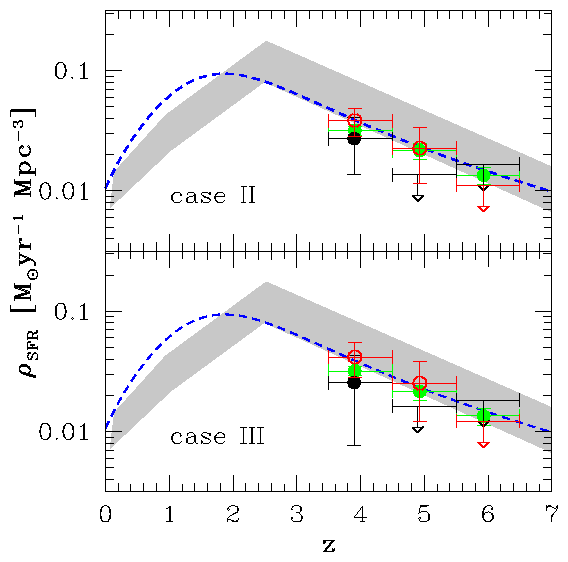}
\caption{SFRD derived for all galaxies (black filled circles) and for the most massive half (red empty circles) in case ii (upper panel) and iii (lower panel), compared with results from \citet[grey region]{Bouwens15} and \citet[dashed blue line]{Madau14}. Such results have been scaled to match the IMF assumed in this work. Green filled points are computed assuming the \citet{Santini15} SFR for the same galaxies included in the stacking analysis.}

\label{fig_sfrd}
\end{figure}

 \noindent The SFR and SFRD values for each redshift bin and considered case are reported in Tab.~\ref{tab4}. Fig.~\ref{fig_sfrd} shows the results as a function of redshift for the two cases (ii and iii) in which we derived the SFR from the stacked X-ray emission. In cases iv and v we did not derive the SFR from X-ray data, but we assumed the values reported by \citet[see \chapt{relative_AGN_SF}]{Santini15}. The corresponding SFRD values are shown as green points in Fig.~\ref{fig_sfrd}. We also plot in Fig.~\ref{fig_sfrd} the SFRD reported by \cite{Bouwens15} and \cite{Madau14}, corrected for the different IMFs used in those works. The estimated SFRD values are fairly consistent with the \cite{Bouwens15} and \cite{Madau14} results. 
 We note that both \cite{Bouwens15} and \cite{Madau14} derived the SFRD at high redshift by integrating UV luminosity functions of galaxies down to magnitudes $\sim1\,\rmn{mag}$ fainter than the limiting magnitude used in this work. However, the brighter magnitude limit of our sample produces only a small underestimate (few per cent) of our SFRD estimates.

 \begin{table*}
%\centering
\caption{Star formation rate [$\rmn{M}_\odot\,\mathrm{yr^{-1}}$] and star formation rate density in units of $10^{-2}\,\rmn{M_\odot yr^{-1} Mpc^{-3}}$ for each stacked sample.}\label{tab4}
\begin{tabular}{|r|r|r|r|r|r|r|r|}
\hline
  \multicolumn{1}{|c|}{{\bf bin}} &
      \multicolumn{1}{|c|}{{\bf $M_*/M_\odot$}} &
  \multicolumn{2}{|c|}{{\bf case ii}} &  
  \multicolumn{2}{|c|}{{\bf  case iii}} &  
  \multicolumn{2}{|c|}{{\bf case iv\&v}}\\  
  
  & &SFR & SFRD & SFR & SFRD& SFR & SFRD\\

  \hline

$3.5\le z< 4.5$ & all                  &$8.7\pm4.3$   &$2.7\pm1.3$ &   $8.2\pm5.7$   & $2.5\pm1.8$ & $10.2\pm0.9$ &  $3.2\pm0.3$\\
$3.5\le z< 4.5$ & $\geq1.32\times10^9$ &$26.6\pm7.1$   &$3.8\pm1.0$ &   $28.8\pm9.4$ & $4.1\pm1.4$ & $15.9\pm1.8$ &  $2.3\pm0.3$\\
$4.5\le z< 5.5$ & all                  & $<11.9$       &$<1.4$      &   $<14.2$      & $<1.6$      & $18.9\pm2.9$ &  $2.2\pm0.3$\\
$4.5\le z< 5.5$ & $\geq1.86\times10^9$ &$42.1\pm20.2$  &$2.2\pm1.1$ &  $46.8\pm24.5$ & $2.5\pm1.3$      & $30.8\pm5.8$ &  $1.7\pm0.3$\\
$5.5\le z< 6.5$ & all                  & $<25.4$       &$<1.7$      &  $<27.8$       & $<1.8$      & $22.5\pm3.3$ &  $1.3\pm0.2$\\
$5.5\le z< 6.5$ & $\geq2.63\times10^9$ & $<36.4$       &$<1.1$      &  $<39.8$       & $<1.2$      & $32.8\pm6.1$ &  $1.0\pm0.2$\\

  \hline

\end{tabular}\\
\end{table*}

 \subsection{Faint-end of the Hard X-ray Luminosity Function (HXLF)}\label{lf}
 
We used the stacked X-ray emission to place constraints on the faint end of the AGN luminosity function by considering that the number of AGN with a certain luminosity $N(L)$ is limited by the observed total AGN luminosity ($L_{AGN}^{TOT}$) following

\begin{equation}\label{upper}
 N(L)\times L \leq L_{AGN}^{TOT},
\end{equation}

\noindent where the equality corresponds to the unrealistic case in which all the measured AGN emission is due to AGN with the same luminosity $L$.
We considered the soft-band flux $F_{half}$ corresponding to half of the \chandra\, sky-coverage in the CANDELS area considered in this work (i.e. half of the area is sensitive to fluxes $<F_{half}$). We then derived the corresponding hard-band luminosity $L_{half}$, including the absorption correction discussed in \chapt{obsc_corr}, at the central redshift of each bin (i.e. $z=$ 4, 5, and 6). We assumed that most of the stacked signal is due to sources close to the 7~Ms detection threshold, with luminosities between $\rmn{log}L_{half} - 1\,\mathrm{dex}$ and $\rmn{log}L_{half}$. For simplicity, the mean logarithmic luminosity, $\mathrm{log}L'=\rmn{log}L_{half} - 0.5\,\mathrm{dex}$ was considered representative of the entire bin. The number of AGN with luminosity in this range could then be derived as 

\begin{equation}\label{upper}
 N(L') \leq L_{AGN}^{stack}/L',
\end{equation}
where $L_{AGN}^{stack}$ is the luminosity due to AGN activity as measured by the stacking analysis in the different cases considered in \chapt{relative_AGN_SF} and already used in \chapt{BHAD} to derive the BHAD. Finally, $N(L')$ was used to estimate upper limits to the LF in the relative 1-dex luminosity bin:

   \begin{equation}
  \psi(\mathrm{log}L')\leq\frac{N_{L'}}{V_c^{'CANDELS}} \,\rmn{dex^{-1}\,Mpc^{-3}}.\\%[1.5ex]
 \end{equation}

\begin{figure*} 
\centering
\includegraphics[width=160mm,keepaspectratio]{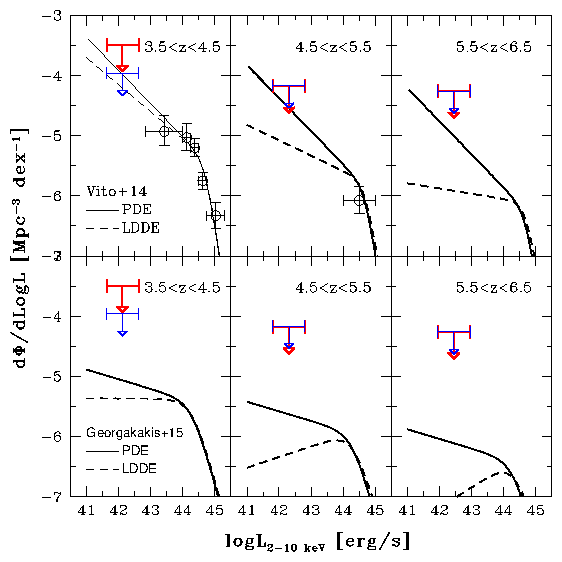}
\caption{Upper limits on the faint end of the AGN luminosity function , for cases i (maximum AGN, red upper limits) and v (AGN+SF, blue upper limits), compared with the PDE and LDDE models of \citet[upper rows]{Vito14} and \citet[lower rows]{Georgakakis15}. The binned luminosity function from \citet{Vito14} computed in the considered redshift bins is also shown for completeness in the upper rows (black points). We do not report the \citet{Georgakakis15} points as they were derived in different redshift bins.}

\label{fig_lf}
\end{figure*}

 In Fig.~\ref{fig_lf} we compare the upper limits to the LF we derived for each redshift bin in case i (all the X-ray emission is due to AGN) and v (X-ray emission due to a mixture of AGN and XRB) with the XLF models of \cite{Vito14} and \cite{Georgakakis15}.
  We have not included any X-ray detected sources; the very small number of possible detected AGN at $z\sim4$ with such low luminosities does not change the results significantly. These results about the faint end of the AGN XLF and their implications are discussed in \chapt{discussion_XLF}.

    \section{Discussion }       
  Using the deepest X-ray data to date, we detected significant ($>3.7\sigma$) stacked X-ray emission from massive $z\sim4$ CANDELS/GOODS-S galaxies (with X-ray detected AGN excluded). Signal with lower-significance ($2.3\sigma$) was also detected from the whole population of $z\sim4$ galaxies (i.e. applying no mass selection). We also reported the first X-ray detection of galaxies at $z\sim5$, with a significance of $2.7\sigma$, corresponding to a $99.7\%$ confidence level. No emission from $z>5.5$ galaxies was significantly detected, even with $\sim10^9\rmn{s}$ effective exposures, several times deeper than previous works \citep[]{Cowie12,Basu-Zych13,Treister13}. These results have been confirmed in Appendix~\ref{LBG} by stacking a sample of LBG from \cite{Bouwens15}. Overall, the signal seems to be sensitive to the stellar mass, with massive galaxies strongly dominating the X-ray emission, while no dependence on the specific star-formation rate properties was derived.
  
  As X-ray emission from galaxies is primarily due to AGN and XRB (which can be used to trace star formation activity), we considered different contributions of these two classes of objects to the measured (or constrained by upper-limits) X-ray emission from our samples (see \chapt{relative_AGN_SF}). Here we discuss the implications of our results for the BHAD, the relative contribution of AGN and XRB (hence star formation) and the faint-end of the XLF. Finally, we apply our stacking procedure to two recently-published samples of high-redshift AGN and compare our findings with results from those works.

\subsection{Black Hole Accretion Rate Density in the Early Universe}  \label{discussion_bhad}
 Besides BH mergers, BH mass growth can happen during the short (compared to a galaxy lifespan) phases of strong accretion (i.e. AGN) and/or through continuous (secular) accretion of material with a low rate during most of the galaxy lifespan \citep[e.g.][]{Hopkins15}, which may not produce enough emission to be detected even in deep X-ray surveys. In Fig.~\ref{fig11} (left panel) our results for AGN and stacked galaxies are compared with models derived from different AGN samples, in order to quantify the relative contribution of inactive galaxies to the BHAD. Even in case i (maximum AGN contribution) the BHAD we derived for the stacked (individually undetected) galaxies is significantly lower than that implied by models of the AGN XLF. This is clearer in cases iv and v, where we have subtracted the contribution from XRBs. Moreover, the BHAD of \textit{detected} AGN in the first redshift bin (the $z=4.5-5.5$ and $z=5.5-6.5$ bins are based on just two and one detections, respectively, and therefore 
are not very informative) is more than one order of magnitude higher than that of undetected galaxies, implying that luminous X-ray detected AGN dominate the BHAD at high redshift, while any potential continuous low-rate accretion in inactive galaxies has a negligible effect on the total BH mass growth. This result agrees with \cite{Volonteri16}, who found that in the Horizon-AGN simulations most of the BH growth at $z>2$ occurs in luminous ($\rmn{log}L_{bol}>43$) AGN. Similarly, at low redshift the \cite{Soltan82} argument and direct measurements ascribe most BH growth to the AGN phase \citep[e.g.][and references therein]{Brandt15}. For completeness, in Fig.~\ref{fig11} (right panel) we compared our results with BHAD predictions from different simulations. Most of them overestimate the BHAD at high redshift, compared to both our results and other observational findings (as per the curves shown in the left panels). Our results suggest that this tension cannot be solved by invoking a low-rate 
accretion in individually undetected galaxies.

The BHAD derived from stacked galaxies would be higher if the fraction of obscured AGN among them we assumed (0.5) were too small. However, even in the extreme case where all the stacked galaxies host heavily obscured AGN (modifying Eq.~\ref{abs} such that all AGN are assumed to be obscured by a column density $\rmn{log}N_H=23.5$), the resulting BHAD would increase only by a factor of $\sim2$, remaining well below the BHAD of X-ray detected sources. Only if all the stacked galaxies hosted Compton-thick AGN the BHAD for galaxies at $z = 3.5 - 4.5$ would be comparable to that of X-ray detected AGN.
 
 The different mixtures of nuclear activity and XRB emission assumed in \chapt{relative_AGN_SF} therefore produce no imprint in the resulting total BHAD, which is due almost entirely to luminous X-ray detected AGN. While at $z\sim4$ the sample size of X-ray detected AGN is statistically meaningful (see Tab.~\ref{tab_Xdet}) and this result is robust, at higher redshift the very limited number of X-ray detected AGN prevents us from drawing solid conclusions. The much smaller number of $z>4.5$ X-ray detected AGN suggests that luminous and rare QSOs, which are missed by deep pencil-beam surveys like the 7 Ms CDF-S, may dominate the BHAD. Indeed, assuming the PDE model of \cite{Vito14} to represent the XLF, AGN with $\mathrm{log}L_X>44$ account for $\sim80\%$ of the BHAD at $4.5<z<5.5$, but, with a comoving space density $\sim10^{-6}\rmn{Mpc^{-3}}$, only $<$1 of such objects is expected to be detected in the area considered in this work. Assuming the \cite{Georgakakis15} model, which has a shallower faint 
end, 
the contribution of luminous AGN to the total BHAD is even larger.

  \subsection{Star formation or nuclear accretion?}\label{SF_or_AGN}
  While the total BHAD, being insensitive to the contribution of X-ray-undetected galaxies, does not give hints on the relative contribution of nuclear accretion and star formation to the stacked X-ray emission, the comparison of our findings in \chapt{SFRD} with independent measurements of the cosmic SFRD available in the literature (Fig.~\ref{fig_sfrd}) is more informative. In particular, if all the emission is assumed to be due to XRB, the SFRD we derived in the three redshift bins is consistent with results from \cite{Bouwens15} and \cite{Madau14}. In other words, unless the $L_X-SFR$ calibrations we used are dramatically wrong, we need X-ray emission from high-redshift galaxies to be largely due to XRB, and hence linked to star formation, in order to match the SFRD derived in other works.  This result agrees with \cite{Cowie12}, who suggested that the upper limits on the X-ray flux they derived at $z\sim6$ by stacking a sample of galaxies are consistent with being entirely related to star formation.
  
  Another hint in favour of an XRB origin of the X-ray emission from individually undetected galaxies can be derived by comparing the SFRD obtained assuming the SFRs from the SED fitting technique by \cite{Santini15} to the SFRD we derived from our stacking analysis. Assuming the SFR by \cite{Santini15} for the same galaxies we included in our stacked samples, Eq.~\ref{SFRD_eq} returns a SFRD values (green points in Fig.~\ref{fig_sfrd}) consistent with what we found from the X-ray stacking analysis only if we consider all the X-ray emission to be due to XRB.
  Stacking the X-ray emission in the hard band did not return detections or upper-limits tight enough to constrain the average X-ray effective spectral slope, which could be used to infer a nuclear or XRB origin of the X-ray emission.

  \subsection{The faint-end of the AGN XLF at high-redshift}\label{discussion_XLF}
  
  The \cite{Vito14} XLF models are marginally consistent with the upper limits resulting from our stacking analysis in both the considered cases (max AGN and AGN+SF) in Fig.~\ref{fig_lf}. We note that these particular models were already considered as upper limits for the AGN XLF, since a maximum redshift-incompleteness correction was applied. Also, \cite{Georgakakis15} showed that assuming the nominal photometric redshifts without considering the associated uncertainties probably led to an overestimate of the AGN space density at $3<z\lesssim5$ in \cite{Vito14}. The models of \cite{Georgakakis15} are largely consistent with the stacked X-ray emission. Those models, while taking into account the Probability Distribution Functions (PDF) of the photometric redshifts of individual AGN, are not corrected for obscuration (i.e. the $x$-axis is the \textit{observed} luminosity). In this respect, they should be considered lower limits for the AGN XLF, as obscured AGN can be detected over smaller areas compared to 
unobscured sources with the same intrinsic luminosity (this, together with the different treatment of the photometric redshifts, could be reflected in the overall lower normalization of their models with respect to the \citealt{Vito14} ones).

  We conclude that our stacking analysis supports a fairly flat faint end of the high-redshift AGN LF, in agreement with previous works in similar redshift ranges \citep{Vito14,Georgakakis15}, but extended to lower luminosities, thanks to the deep \mbox{7 Ms CDF-S} data. Although not highly constraining, the presented results are the first observational estimates of the faint-end of the X-ray-selected AGN LF at high redshift. Still, we cannot distinguish between a PDE- and a LDDE-like evolution at $z>3.5$, as different authors reported quite different slopes and normalizations for the same classes of models.

 Finally, we note that a flat faint end at high redshift can be qualitatively associated with massive BH seeds \cite[e.g. see the discussion in][]{Gilli10}. If SMBH primarily grew from light seeds ($M_{BH}\sim10^2\,M_\odot$), a large number of high-redshift AGN fuelled by accretion onto low-mass SMBH are expected. As the Eddington luminosity scales with the BH mass, even SMBH with accretion rates close to the Eddington limit would populate the faint-end of the LF at high redshift, which would therefore appear steep, as only a small fraction of the SMBH population could reach $\sim L_*$ luminosities. Conversely, if the primary seeding mechanisms involved heavy seeds ($M_{BH}\sim10^{4-5}\,M_\odot$), AGN can more easily reach luminosities close to $L_*$, producing a flatter LF faint end. Future X-ray missions (see \chapt{conclusions}) will provide a better sampling of the AGN XLF at low luminosities and higher redshifts, which is mandatory to understand the mechanisms of 
formation and growth of SMBH in the early universe. If the AGN XLF faint end is found to be flat at higher redshift (i.e. if the space density of low-luminosity AGN declines with increasing redshift at the same rate as that found, e.g., by \citealt{Vito14}, for $L_*$ AGN, or even faster as suggested by \citealt{Georgakakis15}), this would be a strong hint in favour of a massive-seeds scenario.

\subsection{Comparison with previous recent works}
Recently, \cite{Giallongo15} and \cite{Cappelluti16} exploited specialized techniques to detect faint X-ray sources in the 4~Ms CDF-S observations, using the positions of CANDELS-detected galaxies as priors, thereby pushing the {\it Chandra} detection limit to lower fluxes with respect to blind-detection runs. \cite{Giallongo15} selected a sample of 22 $z>4$ AGN candidates by looking for clustering of X-ray counts in space-time-energy parameter space. Five of their AGN  candidates have spectroscopic redshifts, while the remaining ones are selected on the basis of the CANDELS photometric redshifts. Eight out of their 22 sources were already detected in the 4 Ms CDFS catalog of \cite{Xue11}. \cite{Cappelluti16} detected 698 X-ray sources by performing a Maximum Likelihood PSF fit at the positions of CANDELS galaxies simultaneously in the $0.5-2\,\rmn{keV}$ and $2-7\,\rmn{keV}$ bands. Focusing on the redshift range probed by this work, $3.5\leq z<6.5$, and using the CANDELS redshifts, there are 15 objects 
in their catalog, 9 of which were already included in the 4 Ms CDF-S catalog of \cite{Xue11}. Tab.~\ref{Giallongo_Cappelluti} summarizes these numbers.

\begin{table}
%\centering
\caption{Summary of the numbers of sources in the redshift range $3.5\leq z<6.5$ from \citet{Giallongo15} and \citet{Cappelluti16}.}  \label{Giallongo_Cappelluti}
\begin{tabular}{rrrrr}

  \multicolumn{1}{c}{{Sample}} &
 \multicolumn{1}{c}{$N_{\rmn{TOT}}$} &
  \multicolumn{1}{c}{{$N_{\rmn{X11}}$}} &
 \multicolumn{1}{c}{$N_{\rmn{excl}}$} &
 \multicolumn{1}{c}{$N_{\rmn{stack}}$}\\ 
(1)&(2)&(3)&(4)&(5)\\
\hline
G+15 & 22 & 8 & 1 & 13\\
C+16 & 15 & 9 & 1 & 5 \\
\hline
\end{tabular}\\
(1) Parent sample; (2) number of high-redshift AGN candidates in the parent sample; (3) number of sources already detected in the 4 Ms CDF-S catalog of \cite{Xue11}; (4) number of AGN candidates excluded from the stacking because of the proximity of an unrelated X-ray detected source in the 7 Ms CDF-S catalog of Luo et al. (in prep.); (5) number of stacked AGN candidates.
\end{table}

\begin{figure} 
\centering
\includegraphics[width=80mm,keepaspectratio]{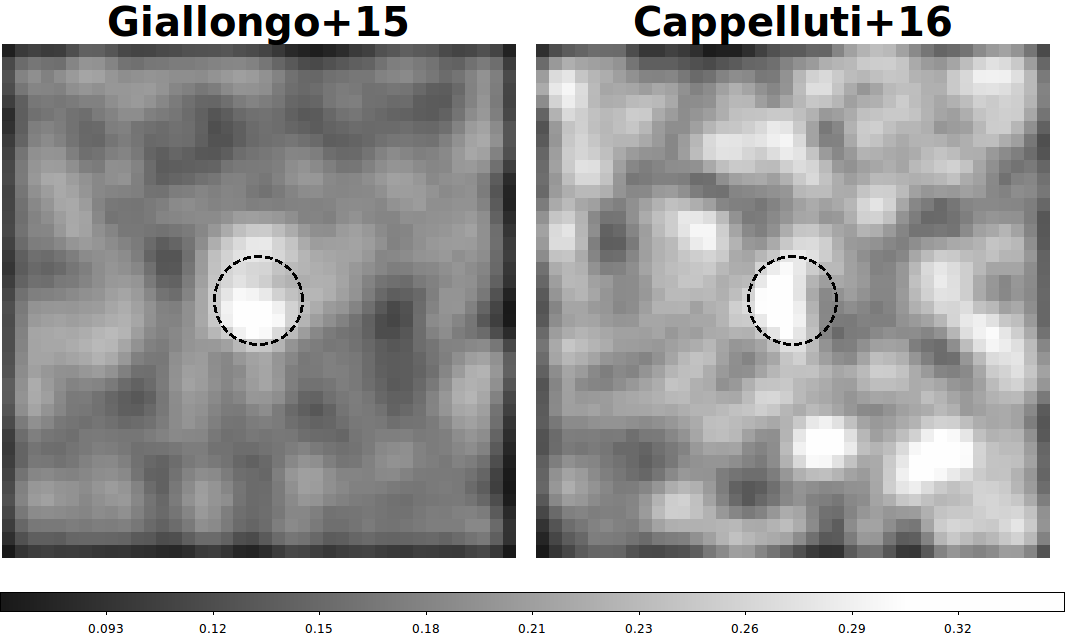}
\caption{X-ray stacked 40x40 pixel images of the 13 \citet{Giallongo15} and 5 \citet{Cappelluti16} sources at $3.5\leq z<6.5$ in the observed $0.5-2\,\rmn{keV}$ band, normalized and smoothed as in Fig.~\ref{fig9}}
\label{stack_Giallongo}
\end{figure}

Applying our stacking code, we derived a total net count rate of $(2.15\pm0.66)\times10^{-5}$ and $(3.36\pm1.99)\times10^{-6}\rmn{cts\,s^{-1}}$ for the  \cite{Giallongo15} and \cite{Cappelluti16} galaxies in the redshift range probed by this work ($3.5\leq z<6.5$), corresponding to a SNR=3.28 and 1.69, respectively. We note that these numbers refer to the stacked galaxies only, which are 13\footnote{CANDELS ID 4285, 5501, 8687, 8884, 9713, 9945, 11287, 12130, 14800, 23757, 28476, 31334, 33073} and 5\footnote{CANDELS ID 4466, 4760,  14537, 24833, 25825} for the two cases, the remaining ones being excluded from the stacking analysis as they are X-ray detected in the Luo et al. (in prep.) catalog or are too close to an X-ray detected source (see Tab.~\ref{Giallongo_Cappelluti}). Fig.~\ref{stack_Giallongo} shows the stacked images in the soft band of the two samples. The numbers reported in Tab.~\ref{Giallongo_Cappelluti} and the stacked signal visible in Fig.~\ref{stack_Giallongo} mean that using the 
CANDELS positions as priors allows the detection of sources undetected by blind detection runs on even deeper (7 Ms) data sets.

Assuming the conversions discussed in \chapt{results} and that all the emission is due to AGN, the stacked count rate for the \cite{Giallongo15} sample corresponds to a total flux of $F_{0.5-2\unit{keV}}=(1.34\pm0.41)\times10^{-16}\unit{erg\,cm^{-2\,}s^{-1}}$ and a luminosity of $L_{2-10\unit{keV}}=(5.22\pm1.59)\times10^{43}\,\unit{erg\,s^{-1}}$ (at $z=4.73$, the median redshift of the sample), a factor $\sim2$ lower than the sum of the luminosities estimated by \cite{Giallongo15} for the considered 13 sources. The stacked count rate of the \cite{Cappelluti16} sample of 5 galaxies corresponds to a total flux of $F_{0.5-2\unit{keV}}=(2.10\pm1.24)\times10^{-17}\unit{erg\,cm^{-2\,}s^{-1}}$ and an AGN total luminosity of $L_{2-10\unit{keV}}=(7.1\pm4.2)\times10^{42}\,\unit{erg\,s^{-1}}$, at the median redshift of the sample, i.e. $z=4.5$.
As for the \cite{Giallongo15} sample, the flux we derive is lower (by a factor of $\sim7$) than that estimated by \cite{Cappelluti16} for the same 5 galaxies, i.e. $\sim1.5\times10^{-16}\unit{erg\,cm^{-2\,}s^{-1}}$. We ascribe the difference of the signal significance between the two samples to the difference in size, as the distributions of net-count rates of the individual sources are similar. Assuming the stellar mass and SFR values from \cite{Santini15} and the \cite{Fragos13} relation between SFR and $L_X$ (as we did in \chapt{results}), we estimated that $\lesssim 10\%$ of the luminosities measured for the two subsamples is due to XRBs.

The discrepancies between the fluxes we derived from our stacking analysis and those reported by \cite{Giallongo15} and \cite{Cappelluti16} for the same objects can be explained by a combination of several factors. For instance, the different methods and assumptions used; e.g. \cite{Cappelluti16} assume a different spectral shape than the one we used. Moreover, we note that, by construction, we stacked the faintest X-ray sources from \cite{Giallongo15} and \cite{Cappelluti16}, as the stacking procedure discards sources detected in the 7 Ms CDF-S catalog. In particular, in the \cite{Cappelluti16} sample, this means that we stacked galaxies with the least secure X-ray counterparts. Indeed, 2 out of the 5 stacked X-ray sources have an ambiguous optical/X-ray matching in \cite{Cappelluti16}. For instance, one of them, CANDELS ID 25825, contributes with a negative flux to the stacked emission and has an X-ray counterpart in \cite{Cappelluti16} at an angular distance comparable to the extraction radius we 
used to derive its photometry (even after the statistical offset correction was applied to the \chandra\, PSF in Appendix~\ref{psf}). As we used the optical positions, we are likely missing most of the X-ray photons accounted for in \cite{Cappelluti16}. This is not true for the \cite{Giallongo15} sample, for which the X-ray fluxes are reported at the optical positions. Finally, as these objects are among the faintest ones detected in the CDF-S, their detection process can be affected by Eddington bias: due to flux statistical fluctuations, detection algorithms can detect sources with average flux below the sensitivity limit of a data-set. Increasing the exposure (e.g. from 4 to 7 Ms), the positive and negative flux fluctuations of a source tend to cancel out, and the measured flux tends to the average value, which is now more reliably measurable. This causes the measured fluxes of faint sources to be preferentially higher in shallower data than in the deeper data.   
 We also note that 3 out of the 5 stacked sources are low-likelihood detections in the soft-band in \cite{Cappelluti16}, and therefore their fluxes should be considered as upper limits.

\begin{figure*} 
\centering
\includegraphics[width=160mm,keepaspectratio]{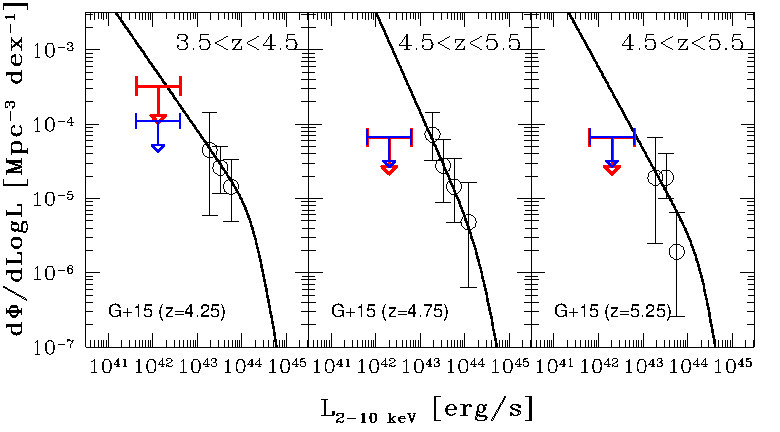}
\caption{Upper limits on the faint end of the AGN luminosity function derived assuming the \citet{Giallongo15} faint-end slope of the AGN LF, transformed into XLF as described in the text, for cases i (maximum AGN, red symbols) and v (AGN+SF, blue symbols). The \citet{Giallongo15} AGN LF model and binned estimates are also reported with black curves and empty circles. The redshifts are slightly different than those of our stacked samples, as shown in the panels.}

\label{lf_Giallongo}
\end{figure*}

\cite{Giallongo15} derived the UV LF of their sample of X-ray detected, high-redshift AGN candidates by deriving the absolute UV magnitude from the apparent optical magnitude in the filter closest to rest-frame $1450\,\angstrom$ at the redshift of each source. Despite the very different approach used, it is interesting to compare the AGN UV LF of \cite{Giallongo15} with the CANDELS stacking results discussed in \chapt{lf}. Thus, we transformed it into an X-ray LF by assuming a SED shape $F_\nu\propto\nu^{-0.5}$ between $1450\,\angstrom$ and $2500\,\angstrom$ (as in \citealt{Georgakakis15}), the \cite{Lusso10} $\alpha_{ox}$, and $\Gamma=1.8$ for the X-ray spectrum. Our upper limits do not agree with the \cite{Giallongo15} AGN luminosity function, especially at $z\sim5$ (see Fig.~\ref{lf_Giallongo}). This could be due to the presence of spurious sources (especially among sources not detected in the 4~Ms CDF-S catalog of \citealt{Xue11}) and/or of low-redshift interlopers in the \cite{Giallongo15} high-redshift 
sample, which would result 
in an overestimate of the slope of the LF faint end and of its normalization. For instance, we note that 7 out of the 8 AGN detected in the \mbox{4~Ms CDF-S} catalog by \cite{Xue11} included in the \cite{Giallongo15} $z>4$ sample, on the basis of the CANDELS photometric redshift \citep{Dahlen13}, have a $z_{phot}<4$ in \cite{Hsu14}, 
which used hybrid (AGN+galaxy) SED templates, with a particular focus on faint X-ray-detected AGN in the 4 Ms CDF-S. These potential issues 
would push the \cite{Giallongo15} AGN LF to lower normalization, in the direction 
of reconciling it with our results. Recently, \cite{Madau15} assumed the \cite{Giallongo15} AGN LF to argue that AGN alone could in principle be responsible for the entire cosmic reionization, under quite strong assumptions, e.g. an escape fraction of hydrogen ionizing photons $f_{esc}=1$. The results discussed above tend to exclude such a possibility, consistent with previous works \citep[e.g.][]{Barger03,Robertson10, Robertson13, Alvarez12, Haardt12,Grissom14}.

\section{Conclusions and future perspectives}\label{conclusions}
In this work, we have investigated the X-ray emission from $3.5\leq z < 6.5$ CANDELS/GOODS-S galaxies by stacking data from the 7 Ms CDF-S, the deepest X-ray survey to date, reaching effective exposures of $>10^9\,\rmn{s}$. We used the results to place constraints on the BHAD and SFRD due to individually undetected galaxies, and on the faint end of the AGN XLF at high redshift. In particular:
\begin{enumerate}
 \item We significantly ($>3.7\sigma$) detected X-ray emission in the $0.5-2\,\rmn{keV}$ band from massive $z\sim4$ galaxies. Signal with lower significance ($2.3\sigma$, $99.0\%$ confidence level) was detected from the entire $z\sim4$ sample (i.e. applying no mass cut).
 Remarkably, we detected a stacked X-ray signal from massive galaxies at $z\sim5$ at a $99.7\%$ confidence level ($2.7\sigma$ ), the highest significance ever obtained for $z\sim5$ galaxies in the X-rays. Overall, X-ray emission is found to be dominated by massive galaxies, while no significant dependence on the sSFR has been found. See \chapt{stacking} and \chapt{mining}.

 \item The BHAD of individually undetected galaxies, derived under reasonable assumptions from the stacked signals or upper limits, is a factor of $\sim10$ lower than that derived by previous works in the literature for the X-ray detected AGN population. This is true even under the least conservative assumption that all the X-ray emission from high-redshift galaxies is due to nuclear accretion (i.e. completely ignoring the contribution from XRB). Notably, the BHAD of a sample of X-ray detected, $z\sim4$ AGN in the CDF-S is in agreement with those works. This means that most of the cosmic BH growth at high redshift occurs in AGN which can be detected in deep X-ray surveys, while continuous, low-rate accretion in X-ray undetected galaxies (which represent by far the majority of the galaxy population) is negligible in the overall BHAD. See \chapt{BHAD} and \chapt{discussion_bhad}.
 
 \item Conversely, assuming that all the X-ray emission is due to XRB and using different scaling relations with the SFR, we found that the SFRD of stacked galaxies is consistent with UV-based results from the literature. Moreover, the total SFR derived from the stacked emission is in agreement with that derived from SED fitting and reported in the \mbox{CANDELS/GOODS-S} catalog. These results strongly suggest that most of the \mbox{X-ray} emission from individually undetected galaxies is due to XRB, and hence to star formation. See \chapt{SFRD} and \chapt{SF_or_AGN}.
 
 \item We presented the first constraints on the faint-end ($\rmn{log}L_X<43$) of the AGN XLF at high redshift. Our findings suggest that the faint-end slope is fairly flat, in agreement with previous works, but extending them down to lower luminosities. We also discussed why improving our knowledge of the evolution of the faint-end slope at high redshifts is important for understanding SMBH seeding and growth processes. See \chapt{lf} and \chapt{discussion_XLF}.
\end{enumerate}

 Improvements of this work and, in general, of our knowledge of high-redshift SMBH and galaxy formation and evolution will take advantage of new optical/IR and X-ray data. In particular, besides the ongoing and future ground-based spectroscopic campaigns in the GOODS-S field, like VUDS \citep{LeFevre15} and VANDELS (http://vandels.inaf.it), which will increase the redshift accuracy and completeness of high-redshift galaxy samples, the upcoming ($\sim$2018) James Webb Space Telescope \citep[\textit{JWST},][]{Gardner06,Gardner09} will provide extremely deep near-to-mid-IR imaging and spectroscopic data, which will open a completely new window on the formation and evolution of galaxies up to $z\sim15-20$ \citep{Finkelstein15}. The study of the first galaxies is indeed among the main goals of the \textit{JWST} mission. While \textit{JWST} will dramatically increase the number of detected low-mass galaxies in the redshift range probed by this work, this will likely not have a strong impact on the results we 
presented, as most (if not all) of the X-ray emission comes from massive galaxies, already identified in existing optical/IR surveys. However, two major improvements due to \textit{JWST} can be foreseen: (1) the galaxy sample selection will be greatly improved by the extremely deep spectroscopic and photometric data provided by \textit{JWST}, resulting in better redshift identifications, more accurate low-redshift interloper removal and thus cleaner stacked samples, and (2) the redshift range of this work will be extended up to $z\sim10$. The last point, in particular, means that we will be able to break into a crucial epoch for our understanding of SMBH and galaxy formation. For instance, sampling the faint-end of the AGN LF at $z>6$ will directly probe SMBH seeding mechanisms in the early universe (see \chapt{discussion_XLF}).

A major step forward in the characterization of the \mbox{X-ray} emission from high-redshift AGN will be taken with the \textit{Athena} \mbox{X-ray} Observatory \citep{Nandra13}, which is expected to detect hundreds of AGN at $z>6$ and, in particular, tens with luminosities as low as $\rmn{log}L_X=43$ \citep{Aird13}. Although \textit{Athena}, which will be launched in $\sim2028$, will not reach the faint fluxes probed by the central region of the \mbox{ 7 Ms CDF-S} (because of the confusion limit imposed by its expected $5''$ angular resolution at very low fluxes), it will survey much larger areas (tens of square degrees) at medium-to-deep sensitivities, thanks to its survey power which is $\sim100$ times faster than that of \textit{Chandra} or \textit{XMM-Newton}. Therefore, while it cannot probe the very faint-end of the AGN XLF, \textit{Athena} will provide valuable new data for studying the evolution of $\lesssim L^*$ AGN up to $z\sim8-10$ and the X-ray properties of rare, powerful quasars perhaps 
at even 
higher redshift. On the basis of the results discussed in this paper (i.e. most of the BHAD at high redshift occurs in moderate-to-high luminosity AGN), X-ray surveys covering wide areas at relatively faint fluxes are best suited to study the black hole accretion history in the early Universe, which is one of the key goals of the \textit{Athena} mission. Valuable wide X-ray surveys would also be provided by the proposed Wide Field X-ray Telescope (WFXT, http://www.wfxt.eu/) mission, which, with its $1\,\rmn{deg^2}$ field of view, $5''$ angular resolution, and collecting area ten times larger than that of \textit{Chandra}, would be a near-perfect survey machine.

The results presented in this paper are relevant to the proposed \mbox{\textit{X-Ray Surveyor}} mission \citep{Weisskopf15} and, in turn, will be greatly improved by the X-ray data it will provide. \mbox{\textit{X-Ray Surveyor}} is the natural successor of \textit{Chandra}, matching its angular resolution, but with a much larger field of view ($10\times$) and sensitivity ($50\times$). These characteristics are reflected in its survey speed, which is a factor of $\sim500$ faster than \textit{Chandra}. The \mbox{\textit{X-Ray Surveyor}} sensitivity will allow sampling of the faint end of the high-redshift AGN XLF and extension of the results of this work. In turn, the evaluation of the number of high-redshift, low-luminosity AGN expected to be detected by \mbox{\textit{X-Ray Surveyor}} must take into account the apparent flatness of the XLF faint end at high redshift, which will limit the low-luminosity source yields. Likely,  \mbox{\textit{X-Ray Surveyor}} will be the first X-ray 
mission which will directly (i.e. by detecting individual objects) sample the very faint-end regime of the AGN XLF at $z>5$ and probe its slope, thus providing us with the needed data (see \chapt{discussion_XLF}), complementary to those provided by \textit{JWST} and \textit{Athena}, to understand the formation and early growth mechanisms of the first SMBH.

\section*{Acknowledgments} 
We thank the anonymous referee for the useful discussion, which improved this work. FV, WNB, and GY acknowledge support from Chandra X-ray Center grant
GO4-15130A and the V.M. Willaman Endowment. FV, RG, CV, and AC acknowledge support from INAF under the contract PRIN-INAF-2014. YQX acknowledges support from the National Thousand Young Talents program, the 973 Program (2015CB857004), NSFC-11473026, NSFC-11421303, the Strategic Priority Research Program of CAS (XDB09000000), and the Fundamental Research Funds for the Central Universities. 

%  \begin{thebibliography}
 \bibliography{biblio}
%   \end{thebibliography}

\appendix
\section{Approximating the \textit{Chandra} PSF with circular regions through accurate simulations}\label{psf}
The aperture photometry procedure described in \chapt{sec1} requires a good knowledge of the PSF fraction enclosed in the radii used to extract the number of counts. However, the \textit{Chandra} PSF is known to have a complex shape, varying over the field of view (f.o.v.). While circular regions are generally good approximations at low off-axis angles ($\theta$), the PSF shape changes dramatically at large $\theta$. Moreover, the PSF is slightly different on different chips and on different positions on the same chip, at a given $\theta$. Knowing the exact PSF shape over the f.o.v. is therefore very difficult. This is why it is generally approximated by a circular region with radius varying with $\theta$. The radial dependence was parametrized differently by different authors \citep[e.g.][]{Hickox06}. In this appendix, we use simulated images of the CDF-S to derive an accurate parametrization of the \textit{Chandra} PSF in the GOODS-S field.

The simulated \textit{Chandra} observations were created with the MARX 5.1.0\footnote{http://space.mit.edu/cxc/marx/} and SAOTRACE 2.0.4\footnote{http://cxc.harvard.edu/cal/Hrma/SAOTrace.html} software. We simulated $\approx2000$ sources at random positions in the CANDELS field with a monochromatic spectrum ($E=1.49\,\rmn{keV}$) and very high fluxes ($F=2\times10^4\rmn{ph\,cm^{-2}\,s^{-1}}$). The chosen energy corresponds to the peak of the \textit{Chandra} efficiency in the soft band. No background was added at this stage. Doing this, we exactly know the detected number of simulated counts for each source, with no statistical issues related to low count rates and background subtraction.\footnote{In \chapt{sec2} we simulated the 4 Ms CDF-S observation in order to validate the stacking procedure. In that case, sources were not simulated with the same high flux but with the distribution of fluxes taken from the \cite{Gilli07} Log$N$-Log$S$ relation and we also included the background obtained by masking all 
the X-ray 
detected sources in the 4 Ms CDF-S.}

As we applied the stacking procedure on the final \mbox{ 7 Ms CDF-S} image, resulting from merging tens of images, we need to account for the different positions each single source has in different observations. To achieve such a result, each source was simulated once per pointing, reproducing exactly the aim point, roll angle, exposure, date and detector offset of the real observations. We used the MARX internal dithering model and an aspect blur of $0.19''$.  

\begin{figure}
\centering
\includegraphics[width=80mm,keepaspectratio]{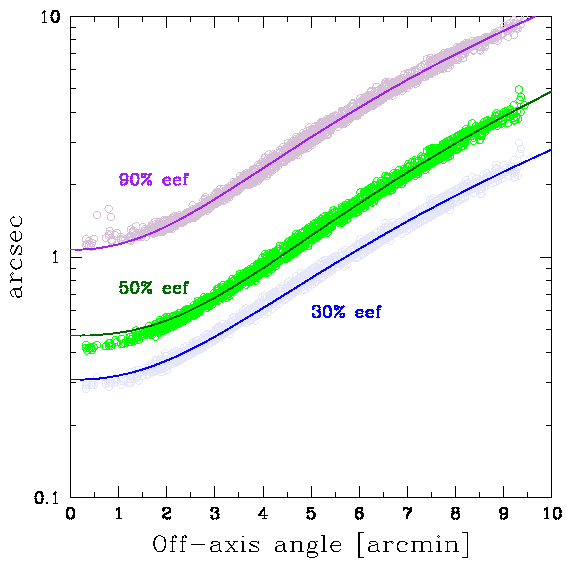}
\includegraphics[width=90mm,keepaspectratio]{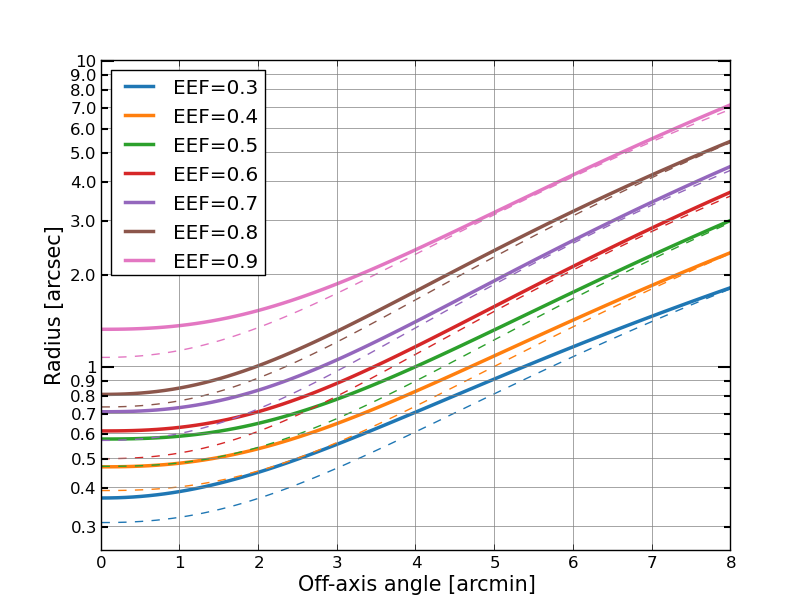}
\caption{{\it Upper panel}: the radii corresponding to the 30, 50, and 90\% EEF for each simulated source are plotted as a function of $\theta$ together with the fitted functional forms (see Eq.~\ref{R}). {\it Lower panel}: the fitted radii corresponding to EEF (from 30\% to 90\% in steps of 10\%) are shown as a function of $\theta$ centring the extraction regions on the X-ray positions (dashed lines) and accounting for the positional scatter of the optical counterparts (solid lines).}

%\caption{{\it Upper panel}: output counts versus input (i.e. 100\% EEF) counts for each individual sources. The line is the locus of the 1:2 relation}
\label{fig2}
\end{figure}

\begin{figure}
\centering
\includegraphics[width=80mm,keepaspectratio]{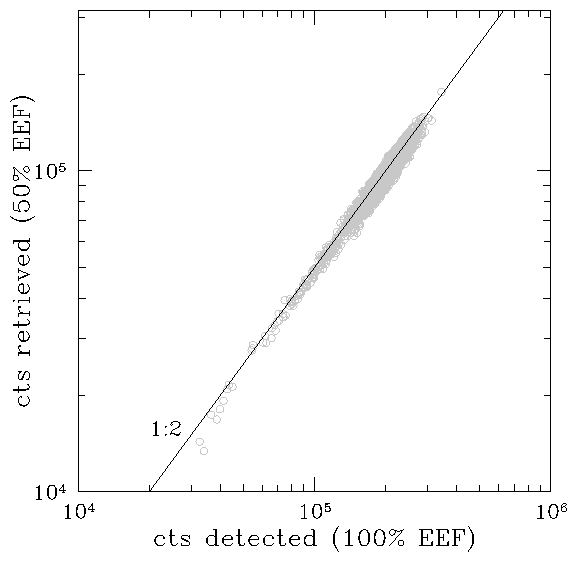}
\caption{Counts retrieved by the aperture photometry procedure inside the 50\% EEF radii versus total (i.e. 100\% EEF) detected counts for each individual simulated source. The line is the locus of the 1:2 relation.}
\label{fig3}
\end{figure}

The simulated images are then reprojected and merged by the CIAO {\it reproject\_obs} tool. 
Through aperture photometry 
we derived the radii of the circular regions which include 30\% to 90\%, in steps of 5\%, of the total simulated photons in the merged image. Simulating $\approx2000$ sources we homogeneously covered the CANDELS field. We then fitted the radius corresponding to each EEF as a function of the off-axis angle ($\theta$, in arcmin), with the function (following \citealt{Hickox06}):
\begin{equation}\label{R}
 R=R_0 + R_{10}\times(\theta/10)^\alpha
\end{equation}
where $R_0$ and $R_{10}$, the radius corresponding to the desired EEF at $\theta=0'$ and $10'$, respectively, are free parameters, as well as the index $\alpha$.  The best-fit parameters are reported in Tab.~\ref{tab1}.
 In Fig.~\ref{fig2}, upper panel, we show the analytical function fitted to the individual radii corresponding to the 30, 50, and 90\% EEF and, lower panel, the resulting best-fit radii as a function of $\theta$ for the considered EEF (in steps of 10\% for clarity). Fig.~\ref{fig3} shows the counts included in the 50\% EEF radius for individual, simulated sources plotted against the total simulated (i.e. 100\% EEF) counts, demonstrating the accuracy of both the best-fitting analytical form and the aperture-photometry procedure. Given that the roll-angles vary for each pointing, each source can be located in more than one chip in different observations. However, each source, depending on the position on the sky, is generally located preferentially in 1--2 chips. For simplicity, we fitted the average behaviour across the chips.
 
 The curves of constant EEF as a function of $\theta$ were derived under the assumption that the circular extraction regions are centred on the X-ray positions. However, in \chapt{sec3} we stacked the CANDELS \textit{optical} positions. Considering the catalog of X-ray detected sources and relative CANDELS counterparts, although \textit{on average} there is no significant offset between the optical and X-ray positions, the offset scattering at low off-axis angles is comparable to the PSF size (see \chapt{stacking}). This causes the EEF at a given $\theta$ corresponding to a given extraction radius to be underestimated, as the extraction region is not centred on the X-ray source. In order to take this effect into account, we fitted again the simulated sources, statistically accounting for the scattering of the optical-X-ray position offsets. In particular, for each simulated X-ray source, we derived the aperture photometry 25 times, each time centring the extraction regions at a distance randomly drawn 
from a 
Gaussian distribution with $\sigma$ equal to the observed offset scattering. The resulting parameters and curves are reported in Tab.~\ref{tab1} and shown in Fig.~\ref{fig2} (lower panel) as solid curves. As expected, accounting for the positional offset scatter causes the radii corresponding to a given EEF to be larger with respect to the case in which no offset is considered (i.e. centring the extraction regions at the X-ray positions). The difference is larger at small $\theta$, where the positional-offset scatter is comparable to the extraction-region sizes, while at large $\theta$ the curves, where the offset scatter is negligible with respect to the extraction-region sizes, converge. By artificially including a broadening due to the offset scattering, we derive an ``effective'' PSF (continuous lines in Fig.~\ref{fig2}) which should not be confused with the true PSF (dashed lines).

\begin{table}
%\centering
\caption{ Fitted parameters of the radii corresponding to different EEF, centring the extraction regions on the X-ray positions (``no offset" case) and accounting for the positional scatter of the optical counterparts (``offset" case). See Eq.~\ref{R}}\label{tab1}.
\begin{tabular}{|r|r|r|r|rrrr}
\hline
  \multicolumn{1}{|c|}{ EEF} &
    \multicolumn{3}{|c|}{{\bf no offset}} &
        \multicolumn{1}{|c|}{} &
    \multicolumn{3}{|c|}{{\bf  offset}} \\
\cline{2-4}\cline{6-8}
    \multicolumn{1}{|c|}{} &
  \multicolumn{1}{|c|}{{\bf $\rmn{R_0}\,\rmn{['']}$ }} &
    \multicolumn{1}{c|}{{\bf $\rmn{R10}\,\rmn{['']}$}} &
  \multicolumn{1}{c|}{{\bf $\rmn{\alpha}$}} &
     \multicolumn{1}{|c|}{} &
  \multicolumn{1}{|c|}{{\bf $\rmn{R_0}\,\rmn{['']}$ }} &
    \multicolumn{1}{c|}{{\bf $\rmn{R10}\,\rmn{['']}$}} &
  \multicolumn{1}{c|}{{\bf $\rmn{\alpha}$}} \\
  \hline
0.30           & 0.31 & 2.49 &  2.29    &    & 0.37 & 2.30 &  2.09 \\
0.35           & 0.35 & 2.91 &  2.36    &    & 0.40 & 2.76 &  2.21 \\
0.40           & 0.39 & 3.40 &  2.48    &    & 0.47 & 3.22 &  2.39 \\
0.45           & 0.42 & 3.87 &  2.50    &    & 0.50 & 3.83 &  2.47 \\
0.50           & 0.47 & 4.40 &  2.54    &    & 0.58 & 4.27 &  2.52 \\
0.55           & 0.50 & 4.93 &  2.54    &    & 0.60 & 4.75 &  2.50 \\
0.60           & 0.50 & 5.31 &  2.38    &    & 0.62 & 5.44 &  2.50  \\
0.65           & 0.55 & 5.90 &  2.40    &    & 0.66 & 5.93 &  2.45 \\
0.70           & 0.57 & 6.39 &  2.31    &    & 0.71 & 6.59 &  2.46 \\
0.75           & 0.64 & 7.07 &  2.32    &    & 0.72 & 7.13 &  2.35 \\
0.80           & 0.74 & 7.89 &  2.34    &    & 0.81 & 7.74 &  2.28 \\
0.85           & 0.82 & 8.55 &  2.24    &    & 0.95 & 8.76 &  2.30 \\
0.90           & 1.07 & 9.65 &  2.22    &    & 1.32 & 10.1 &  2.42 \\

  \hline

\end{tabular}\\

\end{table}

 \section{Assessing the photometric-redshift accuracy}\label{dz}

To quantify the photometric-redshift accuracy, we computed the quantity

\begin{equation}
 \frac{|\Delta z|}{(1+z)} = \frac{|z_{phot}-z_{spec}|}{(1+z_{spec})}
\end{equation}

\noindent for all X-ray-undetected galaxies with a high-quality spectroscopic redshift in the \cite{Santini15} catalog. Considering galaxies at $3.5\leq z<6.5$, the median $\frac{|\Delta z|}{(1+z)}$ is 0.013, with first and third quartiles (Q1 and Q3; corresponding to 25\% and 75\% of the ordered values) of 0.004 and 0.021, respectively. Considering all galaxies with spectroscopic redshifts (i.e. applying no redshift cut), the median $ \frac{|\Delta z|}{(1+z)}$ is 0.020 with Q1 and Q3 equal to 0.009 and 0.038, respectively.

\begin{figure}
\includegraphics[width=80mm,keepaspectratio]{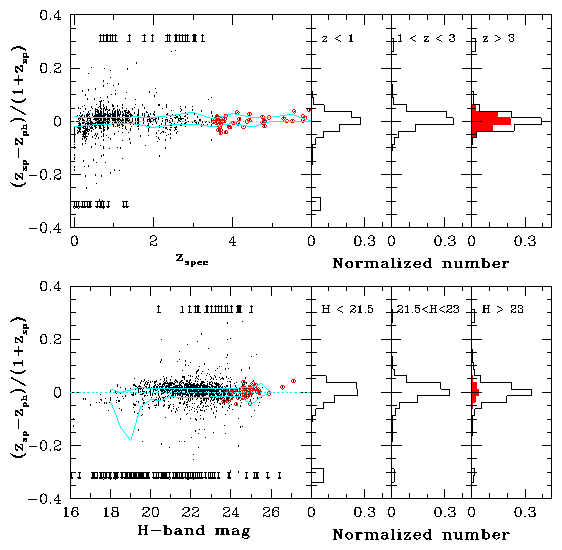}
\caption{Photometric redshift accuracy as a function of redshift (upper panel) and magnitude (lower panel) for all X-ray-undetected galaxies with a high-quality spectroscopic redshift (black points) and those at $z>3.5$ (red points). Extreme outliers are marked with arrows. The $\sigma_{NMAD}^+$ and $\sigma_{NMAD}^-$ curves, computed in bins of 0.5 mag, are shown as cyan lines. We also report the histograms of the photometric redshift accuracy for individual galaxies in three bins of redshift and magnitude.}
\label{fig_dz}
\end{figure}

In Fig.~\ref{fig_dz} we plot the photometric-redshift accuracy as a function of redshift and magnitude to look for a possible degradation at high redshifts and faint magnitudes, typical of our samples of stacked galaxies. We also plotted the curves corresponding to the normalized median absolute deviation 
\begin{equation}
 \sigma_{NMAD}=1.48\times \rmn{Med}\Bigl(\frac{|\Delta z - \rmn{Med(\Delta z)}|}{1+z_{spec}}\Bigr),
\end{equation}

\noindent where $\Delta z = z_{spec}-z_{phot}$, separately for positive ($\sigma_{NMAD}^+$) and negative ($\sigma_{NMAD}^-$) $\Delta z$. From Fig.~\ref{fig_dz} we conclude there is no apparent decrease in the photometric-redshift accuracy at high redshift and faint magnitudes.

These results testify to the good quality of CANDELS photometric redshifts at $z\geq3.5$. Although we did not find an evident degradation of the photometric redshift accuracy at faint magnitudes, the test was performed by construction on the sub-population of CANDELS galaxies for which a spectroscopic redshift was available; these are typically bright ($H<25$) galaxies. Considering fainter galaxies with no spectroscopic identification, the photometric redshift accuracy is expected to be worse: for this reason we will investigate in \chapt{PDF} the importance of redshift uncertainties for our results.

\section{The impact of photometric-redshift uncertainties on the stacking results}\label{PDF}
For the sake of simplicity, 
throughout this work we used the nominal photometric redshifts (i.e. not considering their errors). 
  Although in Appendix~\ref{dz} we quantitatively demonstrated the reliability of CANDELS photometric redshifts at $3.5\leq z<6.5$ for galaxies with magnitude $H\lesssim26$, even a small number of low-redshift interlopers could in principle affect our results. While the proper treatment of photometric redshift uncertainties would be to weight the contribution of every single galaxy  to the total stacked count-rate in a redshift bin considering its Probability Distribution Function (PDF), the computational cost is very expensive (i.e. tens of thousands of CANDELS galaxies would be stacked for each redshift bin, most of which would contribute to the stacked signal with very little weight, since only a small fraction of their PDF overlaps with the considered redshift ranges). Moreover, CANDELS photometric redshifts are evaluated by weighting 6 different PDFs for each object, derived with different SED-fitting codes and assumptions \citep{Dahlen13}. 
  
\begin{figure} 
\centering
\includegraphics[width=80mm,keepaspectratio]{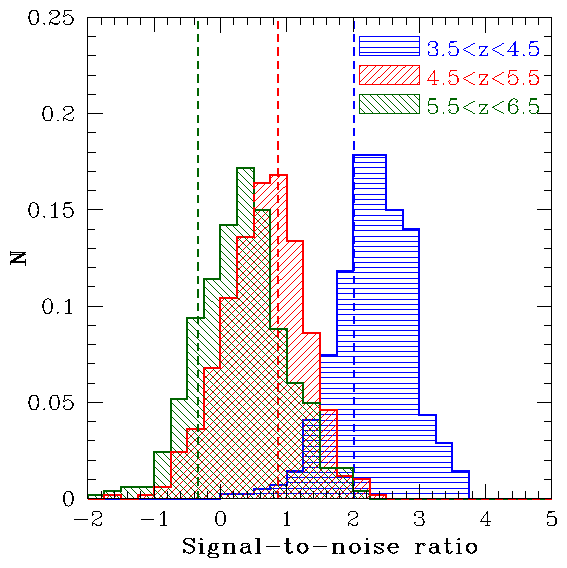}
\caption{Distribution of SNR in the three color-coded redshift bins derived with the Monte Carlo procedure described in \chapt{PDF}. The vertical dashed lines mark the SNR of the stacked emission derived in \chapt{stacking} using the nominal values for CANDELS photometric redshifts.  }

\label{PDF_distro}
\end{figure}

  In this section, we use a Monte Carlo procedure to assess how much the stacked count-rates derived in \chapt{stacking} are affected by contamination due to redshift uncertainties. \cite{Santini15} provide the 68\% confidence interval of the photometric redshift of each CANDELS galaxy. We considered all galaxies for which that interval overlaps the redshift range considered in this work ($3.5\leq z<6.5$) and for each of them extracted randomly a redshift from a Gaussian distribution having the nominal redshift as mean and the redshift error as $\sigma$, i.e. we approximated the individual PDFs with Gaussian functions. Spectroscopic redshifts were always assigned to galaxies for which they are available (see \chapt{sec3.1}).
  We then stacked the count rates of galaxies which fall into the redshift bins of interest, according to the redshift assigned by the random extraction, and compute the SNR as we did in \chapt{stacking}. This procedure was repeated 1000 times, returning a distribution of SNR for each redshift bin. Fig.~\ref{PDF_distro} shows the distribution of SNR derived with the Monte Carlo procedure and the SNR measured in \chapt{stacking} in the three redshift bins.
  The measured SNRs are consistent with the values expected accounting for photometric-redshift uncertainties, although there is a slight offset for the highest redshift bin. We conclude that our results are not significantly affected by contamination from objects wrongly placed in the considered redshift bins.
  The same conclusion was reached assuming for each galaxy a flat PDF in the 68\% confidence interval, with zero probability that the real redshift is outside this interval.

  \section{Stacking a sample of Lyman Break selected galaxies}\label{LBG}
  In order to check how much the stacking results for CANDELS galaxies depend on the particular selection approach, we used a different selection and stacked a sample of Lyman-break galaxies (LBG) from \cite{Bouwens15} at $z=4-8$.
    Although the considered CANDELS and LBG samples are not independent (being selected using the same optical/IR observations), it is worth investigating the results obtained using a different selection method.
 As we did for the CANDELS galaxies, we applied a magnitude cut at $H=28$ to the LBG sample in order to keep the incompleteness of the stacked samples low, and considered only LBG within $7.8$ arcmin of the 7 Ms CDF-S average aim point.  
  Tab.~\ref{LBG_tab} reports the stacking results for the samples of LBG. As for CANDELS galaxies, we detected stacked X-ray emission in the soft-band only for the $z\sim4$ sample (with SNR$\sim2.2$, similar to that derived by stacking CANDELS galaxies in the same redshift bin), while no detection is obtained for higher redshift galaxies (see also \citealt{Basu-Zych13}). This reassures us about the robustness of our results against the particular selection approach applied.
  
We note that these are the best constraints on the stacked X-ray emission of galaxies at redshifts up to 8. In fact, the samples of LBG at $z>4$ are the largest ever stacked and the X-ray data used are the deepest available to date. We reach flux limits of $10^{-19}-10^{-18}\rmn{erg\,cm^{-2}s^{-1}}$.

\begin{figure} 
\centering
\includegraphics[width=80mm,keepaspectratio]{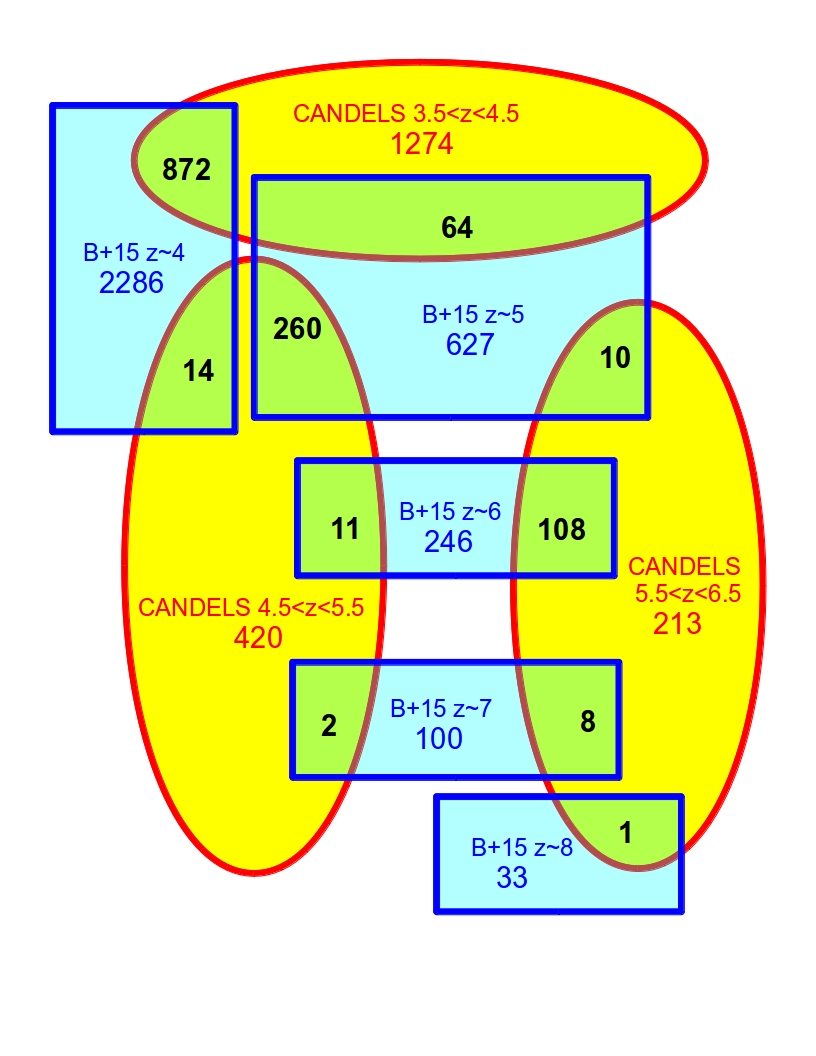}
\caption{Number of stacked galaxies (black numbers in boldface) selected as both CANDELS high-redshift galaxies (yellow ellipses) and LBG by \citet[B+15, blue rectangles]{Bouwens15}, in the different redshift bins. The numbers of stacked galaxies in the parent samples are reported for completeness.}

\label{B15_sample}
\end{figure}

  The overlap between the stacked LBG (3519 galaxies) and CANDELS (2076 galaxies) samples, matched within a $0.2''$ radius, consists of 1350 galaxies.
We show in Fig.~\ref{B15_sample} the overlap among the single, redshift-defined subsamples. We note that \cite{Bouwens15} ran their own detection pipeline and reported only sources which passed the Lyman-break selection. It is therefore difficult to establish if high-redshift CANDELS galaxies not included by \cite{Bouwens15}
did not pass the Lyman-break selection or were not detected at all. We note that the stacked galaxies included in both the CANDELS and LBG samples and those included only in the CANDELS sample have similar magnitude distributions, while those included only in the LBG sample are typically fainter (by $\sim0.5-1\,\rmn{mag}$). Moreover, the individual redshift bins defined for LBG  in \cite{Bouwens15} overlap (e.g. fig.1 in \citealt{Bouwens15} shows that an LBG at $z\sim5.5$ has roughly the same probability of being included in the $z\sim5$ or $z\sim6$ sample, but it also has a small probability to be included in the $z\sim7$ sample; this is due to the Lyman-break method itself) and do not correspond exactly to the ones we defined.

  \begin{table*} 
%\centering
\caption{Main properties of the stacked samples of LBGs.}\label{LBG_tab}

\begin{tabular}{|r|r|r|r|r|r|r|}
\hline
  \multicolumn{1}{|c|}{{ $ z$}} &
  \multicolumn{1}{|c|}{{$N$ }} &
    \multicolumn{1}{|c|}{$\langle H\rangle$} &
    \multicolumn{1}{c|}{{ Exp.}} &
    \multicolumn{1}{c|}{{$CR_{TOT}^w$}} &
     \multicolumn{1}{c|}{{$F^{w,obs,TOT}_{0.5-2\rmn{keV}}$}} &
              \multicolumn{1}{c|}{{$SNR_{boot}$}}\\
              &&mag&($10^9\,\rmn{s}$)&$(10^{-5}\rmn{cts\,s^{-1}})$&($10^{-16}\rmn{erg\,cm^{-2}\,s^{-1}}$)&$\sigma$\\ 
  (1)&(2)&(3)&(4)&(5)&(6)&(7)\\
  \hline
$\sim$4 & 2444 & 26.4 & 14.33 &  $11.30\pm5.12$ & $7.06\pm3.20$& $2.26$ \\
$\sim$5 & 673  & 26.7 & 3.95  & $<2.63$       & $<1.64$         & $0.48$\\
$\sim$6 & 259  & 26.8 & 1.52 & $<1.58$       & $<0.99$          & $-1.89$\\
$\sim$7 & 107  & 27.1 & 0.62 & $<1.03$       & $<0.64$          & $0.60$ \\
$\sim$8 & 36   & 27.1 & 0.21 & $<0.32$       & $<0.20$          & $-1.65$\\

  \hline

\end{tabular}\\
 (1) redshift; (2) number of stacked LBG;  (3) average $H$-band magnitude; (4) total exposure time; (5) total weighted net count-rate and (6) corresponding flux (corrected for the EEF) in the soft band;  (7) signal-to-noise ratio derived from the bootstrap procedure.  
\end{table*}

%   

%   
% 
% \bsp
% 
% 
% 
% 
% \label{lastpage}

\end{document}